\documentclass[a4paper,11pt]{article}
\pdfoutput=1
\usepackage{jheppub}

\usepackage{dsfont}
\usepackage{graphicx}
\usepackage{empheq}
\usepackage{caption}
\usepackage{subcaption}
\usepackage{tikz}
\usetikzlibrary{arrows,shapes}
\usetikzlibrary{trees}
\usetikzlibrary{patterns}
\usetikzlibrary{matrix,arrows} 				
\usetikzlibrary{positioning}				
\usetikzlibrary{calc,through}				
\usetikzlibrary{decorations.pathreplacing}  
\usepackage{pgffor}							

\usetikzlibrary{decorations.pathmorphing}	
\usetikzlibrary{decorations.markings}
\tikzset{
    vector/.style={decorate, decoration={snake}, draw},
    graviton/.style={decorate, double, decoration={snake}, draw},
	provector/.style={decorate, decoration={snake,amplitude=2.5pt}, draw},
	antivector/.style={decorate, decoration={snake,amplitude=-2.5pt}, draw},
        smallvector/.style={decorate, decoration={snake,amplitude=1.5pt,post length=0.5mm}, draw},
    fermion/.style={draw=black, postaction={decorate},
        decoration={markings,mark=at position .55 with {\arrow[draw=black]{>}}}},
    fermionbar/.style={draw=black, postaction={decorate},
        decoration={markings,mark=at position .55 with {\arrow[draw=black]{<}}}},
    fermionnoarrow/.style={draw=black},
    gluon/.style={decorate, draw=black,
        decoration={coil,amplitude=4pt, segment length=5pt}},
    scalar/.style={dashed,draw=black, postaction={decorate},
        decoration={markings,mark=at position .55 with {\arrow[draw=black]{>}}}},
    scalarbar/.style={dashed,draw=black, postaction={decorate},
        decoration={markings,mark=at position .55 with {\arrow[draw=black]{<}}}},
    scalarnoarrow/.style={dashed,draw=black},
    electron/.style={draw=black, postaction={decorate},
        decoration={markings,mark=at position .55 with {\arrow[draw=black]{>}}}},
    bigvector/.style={decorate, decoration={snake,amplitude=4pt}, draw},
    arrow/.style={draw=black, postaction={decorate},
        decoration={markings,mark=at position 1 with {\arrow[draw=black]{>}}}},
}

\tikzstyle{block} = [draw, rectangle, 
    minimum height=3em, minimum width=6earticlem]

    \newcommand{\cb}[1]{
\begin{tikzpicture}[#1]
\draw (-0.2ex,0ex) -- (1.2ex,0ex);
\draw (0.8ex,0ex) -- (0.2ex,-1ex);
\draw (-0.2ex,-1ex) -- (1.2ex,-1ex);
\draw (0.8ex,-1ex) -- (0.2ex,0ex);
\end{tikzpicture}
}

\newcommand{\nb}[1]{
\begin{tikzpicture}[#1]
\draw (-0.2ex,0) -- (1.2ex,0);
\draw (0.8ex,0) -- (0.8ex,-1ex);
\draw (-0.2ex,-1ex) -- (1.2ex,-1ex);
\draw (0.2ex,-1ex) -- (0.2ex,0);
\end{tikzpicture}
}

\newcommand{\tri}[1]{
\begin{tikzpicture}[#1]
\draw (-0.2ex,0) -- (1.2ex,0);
\draw (-0.2ex,-1ex) -- (1.2ex,-1ex);
\draw (0.2ex,0) -- (0.5ex,-0.5ex);
\draw (0.8ex,0) -- (0.5ex,-0.5ex);
\draw (0.5ex,-0.5ex) -- (0.5ex,-1ex);
\end{tikzpicture}
}

\newcommand{\triinv}[1]{
\begin{tikzpicture}[#1]
\draw (-0.2ex,0) -- (1.2ex,0);
\draw (-0.2ex,-1ex) -- (1.2ex,-1ex);
\draw (0.5ex,0ex) -- (0.5ex,-0.5ex);
\draw (0.5ex,-0.5ex) -- (0.2ex,-1ex);
\draw (0.5ex,-0.5ex) -- (0.8ex,-1ex);
\end{tikzpicture}
}

\newcommand{\tric}[1]{
\begin{tikzpicture}[#1]
\draw (-0.2ex,0) -- (1.2ex,0);
\draw (0.2ex,0) -- (0.5ex,-1ex);
\draw (0.8ex,0) -- (0.5ex,-1ex);
\draw (-0.2ex,-1ex) -- (1.2ex,-1ex);
\end{tikzpicture}
}

\newcommand{\tricinv}[1]{
\begin{tikzpicture}[#1]
\draw (-0.2ex,0) -- (1.2ex,0);
\draw (0.2ex,-1ex) -- (0.5ex,0ex);
\draw (0.8ex,-1ex) -- (0.5ex,0ex);
\draw (-0.2ex,-1ex) -- (1.2ex,-1ex);
\end{tikzpicture}
}

\usepackage{xcolor}

\begin{document}

\title{Classical Dynamics of Vortex Solitons from Perturbative Scattering Amplitudes}
\author[a]{Callum R. T. Jones,}
\emailAdd{cjones@physics.ucla.edu}
\affiliation[a]{
Mani L. Bhaumik Institute for Theoretical Physics, Department of Physics and Astronomy, University of California Los Angeles, Los Angeles, CA 90095, USA
}

\abstract{We introduce a novel point-particle effective description of ANO vortex solitons in the critical Abelian Higgs Model (AHM) in $d=2+1$ based on the small winding expansion. Identifying the effective vortices with the elementary quanta of a complex scalar field, relativistic vortex-vortex scattering amplitudes are calculated as a diagrammatic, perturbative expansion in the winding number $N$. Making use of powerful techniques recently developed for analyzing the post-Minkowskian two-body problem in general relativity, we efficiently extract the contribution to the loop integrals from the classical potential region, with the resulting velocity expansion subsequently resummed to all orders. The main result of this paper is an analytic expression for the classical, vortex-vortex potential at $\mathcal{O}\left(N^2\right)$, or one-loop, with exact velocity dependence. By truncating the resulting effective Hamiltonian at $\mathcal{O}\left(p^2\right)$ we derive an analytic, perturbative expression for the metric on the 2-vortex moduli space. Finally, the emergence of the critical AHM from the classical limit of the $\mathcal{N}=2$ supersymmetric AHM, and the resulting constraints on the point-particle EFT is described in detail using an on-shell superspace construction for BPS states in $d=2+1$. 
}

\maketitle
\flushbottom


\section{Introduction}
\label{sec:intro}

Vortices and vortex-strings are topological solitons of central importance across a wide variety of physical systems, from superconductors \cite{ABRIKOSOV1957199} to QCD flux tubes \cite{Nielsen:1973cs} to cosmic strings \cite{Kibble:1976sj}. In spite of decades of study, there are very few analytic results concerning the detailed properties of vortices. The isolated static vortex solution is unknown in closed form, even in the BPS limit \cite{deVega:1976xbp,Taubes:1979tm}, and instead must be obtained approximately by numerically solving a difficult non-linear boundary value problem \cite{deVega:1976xbp,Speight:1996px,Ohashi:2015yta}. Almost all quantitative results concerning the dynamical properties of vortex-vortex interactions are obtained from lattice simulations \cite{Moriarty:1988fx,Shellard:1988zx,Myers:1991yh}, with a few exceptions \cite{Samols:1991ne,Speight:1996px}. Numerical results, even very accurate ones, may obscure analytic features of interest; it is therefore of some importance to develop complimentary analytic perturbative methods \cite{Ohashi:2015yta,Penin:2020cxj,Penin:2021xgr}.

In this paper we will describe an effective field theory approach to vortex-vortex interactions based on the small winding approximation introduced in \cite{Ohashi:2015yta}. By defining an analytic continuation from integer winding number $N$ to general real values, we can set up a perturbative expansion around the limit $N\rightarrow 0$. As demonstrated in \cite{Ohashi:2015yta} there is strong numerical evidence that (after a suitable Pad\'{e} resummation) this expansion gives quantitatively accurate results for the physical value $N=1$. As discussed in more detail in Section \ref{sec:scales}, in the small winding limit there is an additional separation of scales:
\begin{equation}
    \text{Compton wavelength} \hspace{5mm}\ll \hspace{5mm}\text{core size}\hspace{5mm} \ll \hspace{5mm}\text{interaction length}.
\end{equation}
In $d=2+1$, vortex solitons behave in this limit like classical point-particles interacting through the exchange of virtual mediators over a finite distance. The resulting, manifestly relativistic, effective description can then be used to calculate a variety of physical observables, such as vortex-vortex scattering amplitudes, using the standard Feynman diagram expansion. We then proceed by expanding in the limit of small relative velocity and construct a further non-relativistic effective description in which the mediator modes are integrated out generating an effective instantaneous potential. The main result of this paper is (\ref{finalpotential1loop}) an explicit analytic expression for the vortex-vortex potential at $\mathcal{O}\left(N^2\right)$ but with exact velocity dependence. All calculations in this paper are made in the so-called critical limit, defined as the boundary between the type-I and type-II regions, in which the static vortex-vortex forces cancel \cite{deVega:1976xbp,Manton:2002wb}.

Some important technical methods used in this paper are drawn from recent exciting progress concerning the post-Minkowskian two-body problem in general relativity using scattering amplitudes \cite{Cheung:2018wkq,Bjerrum-Bohr:2018xdl,Bern:2019crd}. In that problem, compact astrophysical bodies, black holes and neutron stars, are modelled initially as the elementary excitations of a massive quantum field interacting through the exchange of virtual gravitons \cite{Neill:2013wsa}. The loop integrands of the corresponding quantum scattering amplitudes can then be constructed using a combination of powerful, modern on-shell methods including generalized unitarity \cite{Bern:1994zx,Bern:1994cg} and the double-copy \cite{Kawai:1985xq,Bern:2008qj,Bern:2010ue}. Finite size effects \cite{Cheung:2020sdj,Bern:2020uwk} and spin degrees-of-freedom \cite{Vaidya:2014kza,Bern:2020buy} can be incorporated in a systematic way. One of the most important innovations has been an understanding of how to efficiently extract the contributions relevant for classical observables by expanding the amplitudes \textit{before integration} using the method of regions \cite{Beneke:1997zp}. In particular, this allows for the evaluation of the classical part of the scattering amplitude several orders beyond what is possible in the full quantum problem \cite{Bern:2019nnu,Bern:2021dqo,Bern:2021yeh}. In this paper we will adopt this framework, suitably modified for a gapped system in $2+1$ dimensions, to describe the point-particle vortex effective theory.

The outline of this paper is as follows. In Section \ref{sec:vortex} we review some basic properties of the Abelian Higgs model, the so-called critical limit, and the static vortex soliton solution. In Section \ref{sec:REFT} the relativistic point-particle EFT is defined and in Section \ref{sec:probetree} the cubic vortex-mediator interaction Wilson coefficients are calculated by the matching of probe amplitudes. In Section \ref{sec:susy} the constraints of a hidden supersymmetry are determined using an on-shell superspace and found to agree with the proble calculation. In Section \ref{sec:probeloop} a scheme for calculating the non-linear vortex solution from scattering amplitudes is developed and found to be consistent with the known perturbative solution up to one-loop. In Section \ref{sec:NREFTsec} the non-relativistic EFT is described and in Sections \ref{sec:probetree} and \ref{sec:probeloop}, used to calculate the vortex-vortex potential. Subsequently, in Section \ref{sec:moduli} from the potential is obtained a perturbative approximation for the metric on the 2-vortex moduli space. Further technical details on the velocity resummation of soft integrals and the Fourier transforms of certain loop integrals are deferred to Appendices \ref{sec:soft} and \ref{sec:Fourier} respectively. 

\section{Static ANO Vortex}
\label{sec:vortex}

\subsection{Review: Critical Abelian Higgs Model and the ANO Vortex}
\label{sec:AHM}

In the context of this paper the \textit{full theory} or \textit{UV completion}, is the classical Abelian Higgs model (AHM) in $d=2+1$, defined in our conventions by the action
\begin{equation}
    \label{AHMunbroken}
    S_{\text{AHM}}[\phi,A_\mu] = \int \text{d}^3 x \left[-\frac{1}{4}F_{\mu\nu}^2 +|D_\mu \phi|^2 -\frac{\mu^2}{8}\left(|\phi|^2-v^2\right)^2\right],
\end{equation}
where $D_\mu = \partial_\mu + ie A_\mu$. For $v^2>0$ the $U(1)$ gauge symmetry is spontaneously broken, the model develops a mass gap and the corresponding potential falls off exponentially at large distances. Expanding around the degenerate vacuum 
\begin{equation}
    \phi(x) = \left(v+\frac{\sigma(x)}{\sqrt{2}}\right)e^{i\pi(x)/v},
\end{equation}
the physical degrees of freedom and their interactions become manifest in unitary gauge
\begin{align}
    \label{AHMbroken}
    S_{\text{AHM}}\left[\sigma,A_\mu\right] &= \int \text{d}^3 x\left[-\frac{1}{4}F_{\mu\nu} F^{\mu\nu}+e^2 v^2 A_\mu A^\mu+\frac{1}{2}\left(\partial_\mu\sigma\right)^2-\frac{\mu^2 v^2}{4}\sigma^2\right. \nonumber\\
    &\hspace{20mm}\left.+\sqrt{2}e^2v\sigma A_\mu A^\mu+\frac{e^2}{2} \sigma^2 A_\mu A^\mu -\frac{\mu^2 v}{4\sqrt{2}}\sigma^3 - \frac{\mu^2}{32}\sigma^4\right].
\end{align}
In this form we can read off the perturbative spectrum in the broken phase, it consists of a massive vector boson $A_\mu$, a \textit{massive photon}, and a massive real scalar boson $\sigma$, a \textit{Higgs boson}, with corresponding masses
\begin{equation}
    \label{mediatormasses}
    m_\sigma = \frac{v\mu}{\sqrt{2}}, \hspace{10mm} m_A = \sqrt{2} ev.
\end{equation}
In addition to these elementary particle degrees of freedom, the AHM also admits a spectrum of particle-like topological soliton solutions called \textit{Abrikosov-Nielsen-Olesen (ANO) vortices} \cite{Nielsen:1973cs,ABRIKOSOV1957199}. The static vortex solution is defined as a time-independent, finite energy solution of the classical equations of motion. The solutions are labelled by a toplogical charge, the so-called \textit{winding number}, defined as the degree of the map 
\begin{equation}
    \phi_\infty:S_\infty^1 \rightarrow U(1)\cong S^1,
\end{equation}
from spatial infinity to the degenerate vacuum manifold. Equivalently, the ANO vortex is charged under the magnetic 1-form symmetry corresponding to the conserved 2-form current $\tilde{F}_{\mu\nu}$; the winding number $N$ can be defined in a gauge invariant way as an integral of the magnetic flux 
\begin{equation}
    \label{magflux}
    N \equiv -\frac{e}{2\pi}\int \text{d}^2\mathbf{x} \;F_{12}(\mathbf{x}).
\end{equation}
At long-distances, the static force between a pair of identical vortices is attractive in the type-I regime ($\mu<2e$ or $m_\sigma < m_A$) and repulsive in the type-II regime ($\mu>2e$ or $m_\sigma > m_A$). In general, the form of the classical solution and quantitative information about the dynamics of vortex-vortex interactions can only be determined numerically \cite{Moriarty:1988fx,Shellard:1988zx,Myers:1991yh} or analytically to leading asymptotic order \cite{Nielsen:1973cs,Speight:1996px}.

The situation is a little simpler in the so-called \textit{critical} limit corresponding in our conventions to the case $\mu = 2e$ or
\begin{equation}
    \label{mediatormassescritical}
    m_\sigma = m_A \equiv m.
\end{equation}
In this limit the mass of the vortex can be calculated from a functional of the Bogomolny form \cite{Bogomolny:1975de,PhysRevLett.35.760}
\begin{equation}
    \label{massfuncBPS}
    M = \int \text{d}^2 \mathbf{x} \; \left[\frac{1}{4}\left(F_{12}-e\left(|\phi|^2-v^2\right)\right)^2+\frac{1}{2}|D_1\phi +iD_2\phi|^2\right] +\pi  v^2 N,
\end{equation}
with the critical or BPS vortex solution defined as the solution of the first order \textit{BPS equations} 
\begin{equation}
    \label{BPS}
    F_{12} = e\left(|\phi|^2-v^2\right), \hspace{10mm} D_1\phi +iD_2\phi = 0,
\end{equation}
and the mass of the winding number $N$ solution given by 
\begin{equation}
    \label{BPSvortexmass}
    M = \pi v^2 N.
\end{equation}
It can be shown that the BPS equations admit static multi-soliton solutions and therefore the static force between identical, critical vortices exactly vanishes \cite{Taubes:1979tm}.

Many of these simplifying features of the critical AHM can be understood as a consequence of the fact that the model is a consistent truncation of an $\mathcal{N}=2$ supersymmetric version of the AHM \cite{deVega:1976xbp,Edelstein:1993bb}. This perspective and the consequences for our point-particle EFT construction are described in detail in Section \ref{sec:susy}. For the rest of the paper we will restrict our attention to the critical case, leaving the generalization to the non-critical case to future work. 

\subsection{Perturbative Solution of the BPS Equations}
\label{sec:pertsol}

Even in the critical limit an analytic expression for the static vortex solution is unknown in closed form. For our purposes in this paper, it is sufficient to derive a perturbative solution for the BPS equations by a small modification of the approach described in \cite{deVega:1976xbp}. We begin with an Ansatz for the winding number $N$ vortex in temporal gauge 
\begin{equation}
    \label{Ansatz}
    \phi(\mathbf{x}) = \rho(r)e^{iN\theta}, \hspace{5mm} A_i(\mathbf{x}) = \frac{\epsilon_{ij}x_j}{r^2}A(r), \hspace{5mm} A_0(\mathbf{x}) = 0.
\end{equation}
Inserting this Ansatz into the BPS equations (\ref{BPS}) gives a coupled system of first-order differential equations for the profile functions
\begin{align}
    \frac{A'(r)}{r}+e\left(\rho(r)^2-v^2\right) = 0, \hspace{5mm}\rho'(r)+\frac{\rho(r)}{r}\left( eA(r)-N\right) = 0.
\end{align}
We can solve the first of these equations,
\begin{equation}
\label{rhosol}
    \rho(r) =  \left(v^2-\frac{1}{e}\frac{A'(r)}{r}\right)^{1/2},
\end{equation}
where the branch of the square root is fixed by the boundary condition $\rho \rightarrow v$ as $r\rightarrow \infty$. Inserting this into the second BPS equation
\begin{equation}
\label{Adiff}
    A''(r)+\frac{2e A'(r)A(r)}{r}-\left(2N+1\right)\frac{A'(r)}{r}-2 e^2 v^2 A(r)+2 N e v^2 = 0.
\end{equation}
To solve this differential equation we need appropriate boundary conditions. We impose that the solution is smooth near $r\sim 0$, this requires $A(0)=A'(0)=0$. Similarly, asymptotically expanding (\ref{Adiff}) we find that an asymptotically decaying solution is consistent only if $A(r) \rightarrow N/e$ as $r\rightarrow \infty$. Introducing dimensionless variables 
\begin{equation}
    a(\xi)\equiv eA(r) - N, \hspace{5mm} \xi = \sqrt{2} ev r,
\end{equation}
determining the static vortex solution is equivalent to solving the following second-order, non-linear boundary value problem 
\begin{equation}
    \label{vortexeqs}
    \left(\frac{d^2}{d\xi^2}-\frac{1}{\xi}\frac{d}{d\xi}- 1\right)a(\xi) = -\frac{2 a(\xi) }{\xi}\frac{d a(\xi)}{d\xi}, \hspace{10mm} a(0) = -N, \hspace{10mm} a(\xi \rightarrow \infty) = 0.
\end{equation}
To proceed we follow the approach of \cite{deVega:1976xbp} and rewrite this in the form of an integral equation. A Green's function for the linear differential operator appearing on the left-hand-side 
\begin{equation}
    \left(\frac{d^2}{d\xi^2}-\frac{1}{\xi}\frac{d}{d\xi}- 1\right)G(\xi,\xi') = \delta\left(\xi-\xi'\right),
\end{equation}
is found to be
\begin{equation}
    G(\xi,\xi') = 
    \begin{cases}
        -\xi K_1(\xi) I_1(\xi') & \xi' < \xi \\
        -\xi I_1(\xi) K_1(\xi') & \xi' > \xi
    \end{cases},
\end{equation}
where $K_\nu$ and $I_\nu$ are modified Bessel functions. Imposing the boundary conditions in (\ref{vortexeqs}) we derive the following non-linear integral equation 
\begin{align}
    \label{inteq}
    a(\xi) &= -N\xi K_1(\xi) +2\xi K_1(\xi)  \int_0^\xi \text{d}\xi'\; I_1(\xi')\left[\frac{a(\xi') }{\xi'}\frac{da(\xi')}{d\xi'}\right] \nonumber\\
    &\hspace{12mm}+2\xi I_1(\xi)  \int_\xi^\infty \text{d}\xi'\; K_1(\xi')\left[\frac{a(\xi') }{\xi'}\frac{da(\xi')}{d\xi'}\right].
\end{align}
To perturbatively solve this equation we make use of the \textit{small winding expansion} \cite{Ohashi:2015yta}. The idea is to continue the winding number from the physical quantized values $N\in \mathds{Z}$ to arbitrary real values $N\in \mathds{R}$ and perturbatively expand around the limit $N\rightarrow 0$. We will postpone a discussion of the physical interpretation of this expansion to Section \ref{sec:scales}. On the right-hand-side of (\ref{inteq}) we see that the first term is $\mathcal{O}\left(N\right)$, while the remaining terms are quadratic in $a(\xi)$ and therefore $\mathcal{O}\left(N^2\right)$. So by iterating this integral equation we derive a formal series expansion for the solution
\begin{align}
    \label{vortexinteq}
    a(\xi) &= -\left[N+\frac{\pi}{3\sqrt{3}}N^2\right]\xi K_1(\xi) \nonumber\\
    &\hspace{5mm}+2N^2\xi K_1(\xi)  \int_\xi^\infty \text{d}\xi'\; \xi' I_1(\xi')  K_0(\xi' ) K_1(\xi' )-2N^2 \xi I_1(\xi)  \int_\xi^\infty \text{d}\xi'\; \xi' K_0(\xi' ) \left[K_1(\xi' )\right]^2 \nonumber\\
    &\hspace{5mm}+ \mathcal{O}\left(N^3\right).
\end{align}
It is instructive to compare (\ref{vortexinteq}) with the asymptotic expansion around the limit $\xi\rightarrow \infty$ derived in \cite{deVega:1976xbp} which takes the form 
\begin{equation*}
    a(\xi) = -Z_N \xi K_1(\xi) + \mathcal{O}\left(e^{-2\xi}\right),
\end{equation*} 
and extract an analytic perturbative expansion for the coefficient $Z_N$. The first few orders have a simple form
\begin{equation}
    \label{ZNseries}
    Z_N = N + \frac{\pi}{3\sqrt{3}}N^2 + \frac{\pi^2 }{108}N^3 + \mathcal{O}\left(N^4\right).
\end{equation}
The usefulness of this expansion for physical vortices depends on the radius of convergence of the series, in particular whether or not it gives a good approximation for small \textit{integer} values. Remarkably, it was shown in \cite{Ohashi:2015yta} that for $N=1$ the small winding series (\ref{ZNseries}), truncated at $\mathcal{O}\left(N^6\right)$, quantitatively reproduces independent numerical calculations of $Z_1\approx 1.707864...$ with a numerical discrepancy of $\mathcal{O}\left(10^{-4}\right)$. With the use of suitable (global) Pad\'{e} approximants this numerical discrepancy is further reduced to $\mathcal{O}\left(10^{-6}\right)$. 

Inserting (\ref{vortexinteq}) into the Ansatz (\ref{Ansatz}) and transforming to unitary gauge we find the following perturbative expansion of the static vortex solution
\begin{align}
    \label{pertvortexsol}
    \sigma(\mathbf{x}) &= \sqrt{\frac{2M}{\pi N}}\left[-\left(N+\frac{\pi}{3\sqrt{3}}N^2\right) K_0(m r)-\frac{N^2}{2} \left[K_0(m r)\right]^2\right.\nonumber\\
    &\hspace{35mm}+2N^2 K_0(m r) \int_{m
   r}^{\infty } \text{d}\xi\; \xi I_1(\xi) K_0(\xi) K_1(\xi) \nonumber\\
   &\hspace{35mm}+\left.
   2N^2 I_0(m r) \int_{m r}^{\infty } \text{d}\xi\; \xi K_0(\xi) \left[K_1(\xi)\right]^2 \right] + \mathcal{O}\left(N^{5/2}\right) \nonumber\\
    A_i(\mathbf{x}) &=  \frac{1}{m}\sqrt{\frac{2M}{\pi N}}\frac{\epsilon_{ij} x_j}{r^2}\left[ -\left(N+\frac{\pi}{3\sqrt{3}}N^2\right)mr K_1(mr) \right.\nonumber\\
    &\hspace{35mm}+\left. 2N^2 mr K_1(mr)  \int_{mr}^\infty \text{d}\xi\; \xi I_1(\xi)  K_0(\xi ) K_1(\xi ) \right. \nonumber\\
    &\hspace{35mm}\left.-2N^2 mr I_1(mr)  \int_{mr}^\infty \text{d}\xi\; \xi  K_0(\xi ) \left[K_1(\xi )\right]^2\right] + \mathcal{O}\left(N^{5/2}\right).
\end{align}
In Section \ref{sec:probeloop} we will rederive these expressions using Feynman diagrams.

\section{Relativistic Effective Field Theory}
\label{sec:REFT}

\subsection{Physical Scales and the Small Winding Expansion}
\label{sec:scales}

There are three relevant length scales in the problem: the \textit{quantum size} of the vortex given by the Compton wavelength
\begin{equation}
    R_{\text{Compton}} \sim \frac{1}{M},
\end{equation}
the characteristic \textit{interaction length} between the vortices set by the mediator masses 
\begin{equation}
    R_{\text{interaction}} \sim  \frac{1}{m},
\end{equation}
and the classical or \textit{core size} of the vortex soliton\footnote{
In \cite{Ohashi:2015yta} a precise definition of the core size is given as
\begin{equation}
    R_{\text{core}} \equiv \sqrt{2\times\frac{\int \text{d}^2\mathbf{x} |\mathbf{x}|^2\mathcal{H}(\mathbf{x})}{\int \text{d}^2\mathbf{x} \mathcal{H}(\mathbf{x})}},
\end{equation}
where $\mathcal{H}(\mathbf{x})$ is the energy density of the soliton, and the scaling behaviour (\ref{corescaling}) derived in detail.
}
\begin{equation}
\label{corescaling}
    R_{\text{core}} \sim \frac{\sqrt{N}}{m}.
\end{equation}
The $\sqrt{N}$-scaling of the core size is equivalent to the statement that the density of magnetic flux per unit area in the core is roughly constant as a function of $N$ \cite{Penin:2020cxj,Penin:2021xgr}. For physical vortices, $N\in \mathds{Z}$ and there is no parametric separation of scales between the interaction length and the core size. Physically, this is equivalent to the statement that in an effective description, valid at distances $r \gg R_{\text{core}}$ where the vortex appears as a point-particle, there are only contact interactions. Conversely, at the scale $r\sim R_{\text{interaction}}$ where the vortices interact by exchanging mediator particles, they are not well described as point-particles. 

In this paper, following \cite{Ohashi:2015yta}, we will proceed by continuing from quantized integer values of the winding number $N$ to arbitrary real values and expand around the limit $N\rightarrow 0$. In this regime we instead have a hierarchy of scales of the form 
\begin{equation}    
    \label{ineqcoreint}
    R_{\text{core}} \ll R_{\text{interaction}},
\end{equation}
this means we can try and construct an EFT description at the scale $r\sim R_{\text{interaction}}$, where the vortices behave as point-like particles interacting by exchanging virtual mediator particles. Furthermore, we are interested in the regime in which the vortices can be treated as semi-classical solitons rather than elementary quantum particles, this means we are also assuming
\begin{equation}
    \label{ineqcompcore}
    R_{\text{Compton}} \ll R_{\text{core}}.
\end{equation}
Imposing both (\ref{ineqcoreint}) and (\ref{ineqcompcore}) simultaneously means that at the scale $r\sim R_{\text{interaction}}$ the vortices have an effective description as classical point-like particles of mass $M$ interacting by exchanging mediator particles of mass $m$ and we have a double hierarchy of the form
\begin{equation}
    \label{doubleexp}
    \frac{1}{M} \hspace{2mm}\ll \hspace{2mm}\frac{\sqrt{N}}{m}\hspace{2mm} \ll \hspace{2mm}\frac{1}{m} \hspace{2mm} \sim \hspace{2mm} r.
\end{equation}
The systematics of implementing this double expansion to construct an EFT is described in the following section. In practice we implement this expansion in two steps: first expanding in $N\ll 1$, corresponding to the usual Feynman diagrammatic loop expansion, and then subsequently expanding in $\frac{m}{M}\ll 1$ corresponding to a \textit{classical} or $\hbar$-expansion.

\subsection{Definition of the REFT}

The approach taken to constructing an EFT description of vortices in regime (\ref{doubleexp}) is inspired by recent works on the post-Minkowskian expansion of the two-body problem in classical general relativity \cite{Cheung:2018wkq,Bern:2019crd}. We begin by constructing a relativistic field theory in which the classical point-particle vortices are identified as the quanta of a complex scalar field.\footnote{The ``point-particle" effective field theory described in this section is strongly reminiscent of the well established notion of \textit{particle-vortex duality} \cite{Peskin:1977kp, Dasgupta:1981zz, Karch:2016sxi, Seiberg:2016gmd}. The key difference is the additional separation of scales (\ref{ineqcoreint}). In the REFT the vortex field $\Phi$ self-interacts over a finite distance $r \sim m^{-1}$ by exchanging mediator particles, whereas in the dual XY-model description, vortex-vortex interactions correspond to contact terms of the form $|\Phi|^4$ as appropriate for an effective description valid at distances $r\gg m^{-1}$. It would be very interesting to understand the connection between these two descriptions in more detail.} To distinguish this construction from the non-relativistic effective description given in Section \ref{sec:NREFT}, in which we additionally expand in small velocities, we will refer to this as the \textit{relativistic EFT} or REFT. The general form the of the effective action is 
\begin{equation}
    S_{\text{REFT}}\left[\Phi,\sigma,A_\mu\right] = S_{\text{AHM}}\left[\sigma,A_\mu\right] + S_{\text{vortex}}\left[\Phi,\sigma,A_\mu\right].
\end{equation}
It is convenient to rewrite the Lagrangian parameters of the AHM (\ref{AHMbroken}) in terms of the directly relevant mass scales and the winding number 
\begin{equation}
    e = m\sqrt{\frac{\pi N}{2M}}, \hspace{10mm} v =  \sqrt{\frac{M}{\pi N}}, \hspace{10mm} \mu = m\sqrt{\frac{2\pi N}{M}}.
\end{equation}
In terms of which the AHM action takes the form
\begin{align}
    \label{AHMmanifestcounting}
    &S_{\text{AHM}}[\sigma,A_\mu] \nonumber\\
    &= \int \text{d}^3 x \left[-\frac{1}{4}F_{\mu\nu} F^{\mu\nu}+\frac{1}{2}m^2 A_\mu A^\mu+\frac{1}{2}\left(\partial_\mu\sigma\right)^2-\frac{1}{2}m^2\sigma^2+\sqrt{\frac{\pi}{2}}M^{3/2} \left(\frac{m}{M}\right)^2 N^{1/2}\sigma A_\mu A^\mu\right. \nonumber\\
   &\hspace{17mm}\left. -\sqrt{\frac{\pi}{8}}M^{3/2} \left(\frac{m}{M}\right)^2 N^{1/2}\sigma^3 +\frac{\pi}{4}M\left(\frac{m}{M}\right)^2 N \sigma^2 A_\mu A^\mu - \frac{\pi }{16 }M \left(\frac{m}{M}\right)^2 N\sigma^4 \right],
\end{align}
for reference the Feynman rules associated with this action are given in Appendix \ref{sec:Feynman}. The term denoted $S_{\text{vortex}}$ contains the vortex kinetic term 
\begin{equation}
    \label{Svortex2}
    S^{(2)}_{\text{vortex}}[\Phi] = \int \text{d}^3 x \biggr[|\partial_\mu \Phi|^2 -M^2 |\Phi|^2\biggr],
\end{equation}
as well as the interactions between the vortex and the mediator particles. These are \textit{a priori} unknown and must be determined by matching an observable with full theory. In addition to Lorentz invariance, the REFT is invariant under a global $U(1)$ symmetry, $\Phi \rightarrow e^{i\alpha} \Phi$, the corresponding Noether charge is identified with the topological winding number charge in the full theory. We begin with cubic vortex-mediator interactions, which we expect will dominate at large distances. Up to field redefinition and total derivatives there are three possible interactions
\begin{align}
\label{Svortex3}
    S^{(3)}_{\text{vortex}}\left[\sigma,A_\mu,\Phi\right] &= \int \text{d}^3 x \left[g_s M^{3/2}N^{1/2}\sigma |\Phi|^2 \right.\nonumber\\
    &\hspace{18mm} + ig_e M^{1/2}N^{1/2}A_\mu \left(\Phi\partial^\mu \Phi^* - \Phi^* \partial^\mu \Phi\right)
    \nonumber\\
    &\hspace{18mm}+\left. ig_m M^{-1/2}N^{1/2}\left(\frac{m}{M}\right)^{-1}\epsilon^{\mu\nu\rho}F_{\mu\nu}\left(\Phi\partial_\rho \Phi^* - \Phi^* \partial_\rho \Phi\right)\right].
\end{align}
As we will see in Sections \ref{sec:probetree} and \ref{sec:probeloop}, the cubic vortex-mediator interactions completely determine the static structure of the vortex soliton. As a consequence these interactions should be invariant under the same symmetries as the solution (\ref{Ansatz}). We will therefore impose that the cubic interactions are invariant under parity acting as
\begin{align}   
    \label{parity}
    &A_\mu(x) \xrightarrow[]{\text{P}} -{\mathcal{P}_\mu}^\nu A_\nu(\mathcal{P}\cdot x), \hspace{5mm} \sigma(x) \xrightarrow[]{\text{P}} \sigma(\mathcal{P}\cdot x), \hspace{5mm} \Phi(x) \xrightarrow[]{\text{P}} \Phi(\mathcal{P}\cdot x),
\end{align}
 where ${\mathcal{P}_\mu}^\nu \equiv \text{diag}\left(1,-1,1\right)$. This is a symmetry of (\ref{Svortex3}) only if the ``electric" coupling vanishes, $g_e=0$.\footnote{
This interaction can be generated by adding a Chern-Simons term to the AHM, which explicitly breaks parity invariance, and would source the $A_r$ component of the static soliton solution \cite{Paul:1986ix}.}  

In (\ref{Svortex3}) we have chosen a convenient normalization for the Wilson coefficients that ensures uniform scaling of interactions in the soft region defined in Section \ref{sec:NREFT}. Furthermore we have assumed that the cubic vortex-mediator coupling scales as $\sim N^{1/2}$. Combined, with the explicit $N$-scaling of the interactions in (\ref{AHMmanifestcounting}), this ensures that the $N$-expansion aligns with the usual diagrammatic loop expansion. 

We expect that the REFT must also contain higher-multiplicity interactions between the vortex and $n$-mediator fields with $n\geq 2$. These interactions do not contribute to the static structure, rather they encode the response of the finite size soliton to external perturbations, in the gravitational context such Wilson coefficients are usually called Love numbers \cite{Bini:2020flp,Cheung:2020sdj,Bern:2020uwk}. In Section \ref{sec:loop} we will see that certain quartic or ``seagull" interactions must necessarily be non-zero to ensure the expected cancellation of static vortex-vortex forces in the critical limit. Otherwise we will leave a determination of such finite size effects to future work.

Finally, as discussed briefly in Section \ref{sec:AHM} the critical limit of the AHM in $d=2+1$ corresponds to a consistent truncation of a model with $\mathcal{N}=2$ supersymmetry \cite{deVega:1976xbp,Edelstein:1993bb}. Even though this larger model contains additional degrees of freedom that are eventually decoupled, the enlarged symmetry implies non-trivial constraints on the self-interactions between the states that remain. Importantly, this residual supersymmetry should also constrain the, \textit{a priori} unknown, interactions between the mediator fields and the vortices in the point-particle REFT. Especially at higher-multiplicity, this hidden symmetry will greatly reduce the number of Wilson coefficients that need to be matched between full theory and the REFT. In the following subsection we will determine these constraints at cubic order by explicitly constructing the corresponding superamplitudes.

\subsection{Supersymmetry Constraints}
\label{sec:susy}

We begin by constructing an on-shell representation of the $\mathcal{N}=2$ supersymmetry algebra; it is convenient to write this in a form where the two real valued supercharges are combined into a single complex valued supercharge
\begin{equation}    
\label{N2susyalg}
    \{Q_a, Q^\dagger_b\} = 2\sigma_{ab}^\mu P_\mu -2iZ\epsilon_{ab}, \hspace{5mm} \{Q_a,Q_b\} = 0, \hspace{5mm} \{Q_a^\dagger,Q_b^\dagger\} = 0,
\end{equation}
where we have included a central charge $Z$ corresponding to the winding number \cite{Edelstein:1993bb}. We would like to write down a representation of this algebra on the space of (outgoing) one-particle states with momentum $p^\mu$ and mass $m>0$. For this purpose it is convenient to introduce massive spinor variables $\lambda_a$ and $\tilde{\lambda}_a$ defined in Appendix \ref{sec:conventions}, and form the little group covariant supercharges 
\begin{equation}
    q = \frac{1}{\sqrt{2}m}\lambda^a Q_a, \hspace{5mm} \tilde{q} = \frac{1}{\sqrt{2}m}\tilde{\lambda}^a Q_a, \hspace{5mm} q^\dagger = \frac{1}{\sqrt{2}m}\tilde{\lambda}^a Q^\dagger_a, \hspace{5mm} \tilde{q}^\dagger = \frac{1}{\sqrt{2}m}\lambda^a Q_a^\dagger,
\end{equation}
satisfying the simpler algebra
\begin{equation}
    \{q,q^\dagger\} = 1-\frac{Z}{m}, \hspace{10mm} \{\tilde{q},\tilde{q}^\dagger\} = 1+\frac{Z}{m},
\end{equation}
with all other anti-commutators vanishing. In this form it is simple to write down a representation of this algebra as multiplicative and differential operators on the space of Grassmann polynomials in two variables $\eta$ and $\tilde{\eta}$ 
\begin{align}
    &q = \left(1-\frac{Z}{m}\right)^{1/2}\eta, \hspace{5mm} q^\dagger = \left(1-\frac{Z}{m}\right)^{1/2}\frac{\partial}{\partial \eta}, \nonumber\\
    &\tilde{q} = \left(1+\frac{Z}{m}\right)^{1/2}\tilde{\eta}, \hspace{5mm} \tilde{q}^\dagger = \left(1+\frac{Z}{m}\right)^{1/2}\frac{\partial}{\partial \tilde{\eta}}.
\end{align}
Translating back to the original spinor supercharges we have a representation of the superalgebra
\begin{align}
    Q_a &= i\sqrt{2}\left[ \left(1-\frac{Z}{m}\right)^{1/2}\tilde{\lambda}_a\eta -  \left(1+\frac{Z}{m}\right)^{1/2}\lambda_a \tilde{\eta}\right] \nonumber\\
    Q^\dagger_a &= -i\sqrt{2}\left[\left(1-\frac{Z}{m}\right)^{1/2} \lambda_a\frac{\partial}{\partial \eta} -  \left(1+\frac{Z}{m}\right)^{1/2} \tilde{\lambda}_a\frac{\partial}{\partial \tilde{\eta}}\right].
\end{align}
In this form we see that there are two special cases, if $Z=+m$ or $Z=-m$ the dependence on $\eta$ or $\tilde{\eta}$ respectively drops out. These correspond to the BPS and anti-BPS limits in which the supermultiplets are shortened. 

The spectrum of the $\mathcal{N}=2$ AHM (in the broken phase) consists of a pair of long multiplets containing the mediator particles. In this on-shell superspace formalism these correspond to the on-shell superfields\footnote{See \cite{Elvang:2011fx,Agarwal:2013tpa} for a general discussion of on-shell supersymmetry methods for scattering amplitudes with less than maximal supersymmetry.}
\begin{equation}
    \Gamma = \psi^+ + \varphi \eta + \gamma^+ \tilde{\eta} + \tilde{\psi}^+ \eta \tilde{\eta}, \hspace{10mm} \overline{\Gamma} = \psi^- + \gamma^{-} \eta + \overline{\varphi} \tilde{\eta} + \tilde{\psi}^- \eta \tilde{\eta},
\end{equation}
related to each other by a parity transformation, but to themselves under CPT. In this context the central charge $Z$ is identified with the winding number of the vortex solitons, and so for the mediator particles $Z=0$ or 
\begin{align}
    Q_a = i\sqrt{2}\left[ \tilde{\lambda}_a\eta -  \lambda_a \tilde{\eta}\right], \hspace{10mm}
    Q^\dagger_a = -i\sqrt{2}\left[\lambda_a\frac{\partial}{\partial \eta} -  \tilde{\lambda}_a\frac{\partial}{\partial \tilde{\eta}}\right].
\end{align}
It is interesting to compare this spectrum to the non-supersymmetric AHM (\ref{AHMbroken}). In addition to the fermions $\psi^\pm$ and $\tilde{\psi}^\pm$, to form the necessary long supermultiplets we have had to introduce two real scalar degrees of freedom $\varphi$ and $\overline{\varphi}$, while the non-supersymmetric model contains only a single real scalar Higgs boson $\sigma$. It therefore must be the case that in an appropriate classical limit, one of the scalar degrees of freedom in the $\mathcal{N}=2$ AHM consistently decouples. By considering the action of parity it is straightforward to identify 
\begin{equation}
    \sigma = \frac{1}{\sqrt{2}}\left(\varphi + \overline{\varphi}\right), \hspace{10mm} \tilde{\sigma} = \frac{1}{\sqrt{2}}\left(\varphi - \overline{\varphi}\right),
\end{equation}
where the decoupling orthogonal linear combination $\tilde{\sigma}$ is a pseudo-scalar. 

In the $\mathcal{N}=2$ version of the REFT we introduce a BPS and anti-BPS multiplet corresponding to the vortex and anti-vortex respectively \cite{Intriligator:2013lca}. For the vortex multiplet with $Z=M$ the supercharges are
\begin{equation}
    Q_a = -2i \lambda_a \tilde{\eta},\hspace{10mm} Q_a^\dagger = 2i \tilde{\lambda}_a \frac{\partial}{\partial \tilde{\eta}},
\end{equation}
and the on-shell superfield
\begin{equation}
    \label{vortexsuperfield}
    \Sigma = \Phi + \tilde{\eta} \Psi^+.
\end{equation}
For the anti-vortex multiplet with $Z=-M$ the supercharges are 
\begin{equation}
    Q_a = 2i\tilde{\lambda}_a \eta, \hspace{10mm} Q^\dagger_a = -2i\lambda_a \frac{\partial}{\partial \eta},
\end{equation}
and the on-shell superfield 
\begin{equation}
    \label{antivortexsuperfield}
    \overline{\Sigma} = \tilde{\Psi}^+ + \eta\overline{\Phi}.
\end{equation}
From (\ref{vortexsuperfield}) and (\ref{antivortexsuperfield}) we observe that the spin of the fermionic components of both the vortex and anti-vortex superfields are the same. This means that the supersymmetrization of the REFT must violate parity, which changes the sign of the spin. 

\subsubsection*{Cubic mediator interactions}

As a first application we will reproduce the specific relative tuning of the cubic interactions in the critical limit of the AHM (\ref{AHMmanifestcounting}). The most general cubic superamplitudes we can write down have the form
\begin{align}
    \mathcal{M}_3\left(\Gamma_1,\Gamma_2,\Gamma_3\right) &= c_{\Gamma\Gamma\Gamma}\left[\delta^{(2)}\left(Q\right)\right]_{\Gamma\Gamma\Gamma} \times \biggr[[23] \eta_1 +[31] \eta_2 +[12] \eta_3\biggr] \nonumber\\
    \mathcal{M}_3\left(\overline{\Gamma}_1,\Gamma_2,\Gamma_3\right) &= c_{\overline{\Gamma}\Gamma\Gamma}\left[\delta^{(2)}\left(Q\right)\right]_{\Gamma\Gamma\Gamma} \times \biggr[[23] \tilde{\eta}_1+\langle 13] \eta_2 -\langle 12] \eta_3\biggr],
\end{align}
where 
\begin{equation}
    \left[\delta^{(2)}\left(Q\right)\right]_{\Gamma\Gamma\Gamma} \equiv \sum_{i<j}\left(\langle ij\rangle \eta_i \eta_j+[ij] \tilde{\eta}_i \tilde{\eta}_j-\langle ij] \eta_i \tilde{\eta}_j-\langle ji] \eta_j \tilde{\eta}_i \right)-im\sum_{i} \eta_i \tilde{\eta}_i.
\end{equation}
The remaining cubic superamplitudes are assumed to be related to the above by parity 
\begin{align}
\mathcal{M}_3\left(\overline{\Gamma}_1,\overline{\Gamma}_2,\overline{\Gamma}_3\right) &= \mathcal{M}_3\left(\Gamma_1,\Gamma_2,\Gamma_3\right)\biggr\vert_{\lambda \leftrightarrow \tilde{\lambda},\;\eta \leftrightarrow \tilde{\eta}}\nonumber\\
\mathcal{M}_3\left(\Gamma_1,\overline{\Gamma}_2,\overline{\Gamma}_3\right) &= \mathcal{M}_3\left(\overline{\Gamma}_1,\Gamma_2,\Gamma_3\right)\biggr\vert_{\lambda \leftrightarrow \tilde{\lambda},\;\eta \leftrightarrow \tilde{\eta}}.
\end{align}
We find that there is a two-parameter family of parity-invariant cubic interactions, labelled by the independent coefficients $c_{\Gamma\Gamma\Gamma}$ and $c_{\overline{\Gamma}\Gamma\Gamma}$, with the assumed spectrum and $\mathcal{N}=2$ supersymmetry. To recover the AHM we must also impose that the real scalar $\tilde{\sigma}$ consistently decouples. Projecting out the component amplitudes we find a non-trivial condition 
\begin{equation}
    \mathcal{M}_3\left(\tilde{\sigma}_1,\gamma_2^+,\gamma_3^+\right) \propto \mathcal{M}_3\left(\varphi_1,\gamma_2^+,\gamma_3^+\right) - \mathcal{M}_3\left(\overline{\varphi}_1,\gamma_2^+,\gamma_3^+\right) =  \left(c_{\Gamma\Gamma\Gamma} - c_{\overline{\Gamma}\Gamma\Gamma}\right) [23]^2.
\end{equation}
Requiring that this vanishes gives $c_{\Gamma\Gamma\Gamma} = c_{\overline{\Gamma}\Gamma\Gamma}$. It is straightforward to check that all other component amplitudes with a single $\tilde{\sigma}$ vanish automatically. After imposing this condition the non-decoupled bosonic component amplitudes are found to be 
\begin{equation}
    \mathcal{M}_3\left(\sigma_1,\sigma_2,\sigma_3\right) = -\frac{3}{\sqrt{2}}m^2 c_{\Gamma\Gamma\Gamma},\hspace{5mm} \mathcal{M}_3\left(\sigma_1,\gamma_2,\gamma_3\right) = \sqrt{2}m^2 c_{\Gamma\Gamma\Gamma} \left(\epsilon_2\cdot \epsilon_3\right).
\end{equation}
Comparing this with the on-shell 3-particle amplitudes calculated from the critical AHM (\ref{AHMmanifestcounting}) we find that this reproduces exactly the expected relative tuning of coefficients where we identify
\begin{equation}
    c_{\Gamma\Gamma\Gamma} = \sqrt{\pi} M^{-1/2} N^{1/2}.
\end{equation}

\subsubsection*{Cubic vortex-mediator interactions}

Next we repeat the above analysis for cubic interactions between the vortices and mediators. Here we find that there is a two-parameter family of superamplitudes
\begin{align}
    \mathcal{M}_3\left(\Gamma_1,\Sigma_2,\overline{\Sigma}_3\right) &= \left[\delta^{(2)}\left(Q\right)\right]_{\Gamma\Sigma\overline{\Sigma}}\times \left[c_{\Gamma\Sigma\overline{\Sigma}}\;[13\rangle\right], \nonumber\\
    \mathcal{M}_3\left(\overline{\Gamma}_1,\Sigma_2,\overline{\Sigma}_3\right) &= \left[\delta^{(2)}\left(Q\right)\right]_{\Gamma\Sigma\overline{\Sigma}}\times \left[c_{\overline{\Gamma}\Sigma\overline{\Sigma}}\;\langle 13\rangle\right],
\end{align}
where
\begin{equation}
    \left[\delta^{(2)}\left(Q\right)\right]_{\Gamma\Sigma\overline{\Sigma}} \equiv -im\eta_1 \tilde{\eta}_1 + \sqrt{2}\langle 12\rangle \eta_1 \tilde{\eta}_2 -\sqrt{2}\langle 13] \eta_1 \eta_3 - \sqrt{2}[12\rangle \tilde{\eta}_1 \tilde{\eta}_2 +\sqrt{2} [13]\tilde{\eta}_1 \eta_3 +2\langle 23] \tilde{\eta}_2 \eta_3. 
\end{equation}
Projecting out the bosonic component amplitudes we find
\begin{align}
    &\mathcal{M}_3\left(\varphi_1,\Phi_2,\overline{\Phi}_3\right) = \frac{c_{\Gamma\Sigma\overline{\Sigma}}}{2\sqrt{2}}\;m\left(m-2M\right), \hspace{10mm} \mathcal{M}_3\left(\overline{\varphi}_1,\Phi_2,\overline{\Phi}_3\right) = -\frac{c_{\overline{\Gamma}\Sigma\overline{\Sigma}}}{2\sqrt{2}}\;m\left(m+2M\right) \nonumber\\
    &\mathcal{M}_3\left(\gamma^+_1,\Phi_2,\overline{\Phi}_3\right) = -\sqrt{2}c_{\Gamma\Sigma\overline{\Sigma}}\;[13]\langle 31], \hspace{11.5mm} \mathcal{M}_3\left(\gamma^-_1,\Phi_2,\overline{\Phi}_3\right) =\sqrt{2}c_{\overline{\Gamma}\Sigma\overline{\Sigma}}\;\langle 13]\langle 31\rangle.
\end{align}
Here we encounter an interesting puzzle. If we insist that the cubic vortex-photon interaction takes the ``magnetic" form (\ref{Svortex3}) then the couplings must be related as $c_{\Gamma\Sigma\overline{\Sigma}} = c_{\overline{\Gamma}\Sigma\overline{\Sigma}}$. But this choice gives the following values for the cubic vortex-scalar interactions
\begin{equation}    
\label{cubicvortexscalarint}
    \mathcal{M}_3\left(\sigma_1,\Phi_2,\overline{\Phi}_3\right) = -mMc_{\Gamma\Sigma\overline{\Sigma}}, \hspace{10mm} \mathcal{M}_3\left(\tilde{\sigma}_1,\Phi_2,\overline{\Phi}_3\right) = \frac{1}{2}m^2 c_{\Gamma\Sigma\overline{\Sigma}},
\end{equation}
so we find that the pseudoscalar $\tilde{\sigma}$ does not decouple. We do observe however that the $\tilde{\sigma}$ coupling is suppressed by an additional power of $\frac{m}{M}$ relative to the $\sigma$ coupling, and so it will decouple in the classical limit; we will revisit this observation in Section \ref{sec:tree}. Matching with the general form of the cubic interactions in the REFT, this superspace analysis implies the following relation between Wilson coefficients
\begin{equation}
    \label{susypredcubic}
    g_s = 4g_m.
\end{equation}
In Section \ref{sec:probetree} we will verify this prediction by explicitly matching the REFT with the perturbative vortex solution.

\subsection{Matching with Full Theory: Probe Scattering}
\label{sec:probetree}

To determine the Wilson coefficients in the REFT we need to match an observable with full theory. As we will see the cubic vortex-mediator interactions (\ref{Svortex3}) are fixed by matching a probe amplitude in both descriptions. We introduce a fictitious complex scalar probe particle $\chi$ with scalar and magnetic type cubic interactions with the mediator fields
\begin{align}  
\label{Sprobe}
    &S_{\text{Probe}}[\chi,\sigma,A_\mu] =\int \text{d}^3 x \biggr[|\partial_\mu \chi|^2 -\mathfrak{m}^2 |\chi|^2 +\mathfrak{g}_s \sigma|\chi|^2+i\mathfrak{g}_m \epsilon^{\mu\nu\rho}F_{\mu\nu}\left(\chi^* \partial_\rho \chi-\chi \partial_\rho \chi^*\right)\biggr],
\end{align}
where $\mathfrak{g}_{s,m}$ can be taken to be arbitrarily small. To calculate a probe amplitude in full theory we begin by calculating an off-shell two-point function in a vortex background
\begin{equation}
    \langle \chi^*(x_1) \chi(x_2) \rangle_{\overline{\sigma},\overline{A}_\mu} = \int \mathcal{D}\chi \mathcal{D}\sigma \mathcal{D}A_\mu \left[\chi^*(x_1) \chi(x_2)\right] e^{iS_{\text{AHM}}[\overline{\sigma}+\sigma,\overline{A}_\mu + A_\mu]+iS_{\text{Probe}}[\chi,\overline{\sigma}+\sigma,\overline{A}_\mu + A_\mu] },
\end{equation}
with the usual normalization factor implicit. We calculate this two-point function perturbatively in $N$, where the expansion of the background, here denoted $\overline{\sigma}$, $\overline{A}_\mu$, is given by iteratively solving the BPS equations (\ref{pertvortexsol}), working to leading-order in the probe couplings or $\mathcal{O}\left(\mathfrak{g}_{s,m}\right)$. Terms in $S_{\text{AHM}}$ that are zeroth order in the perturbation $\sigma,A_\mu$ (depend only on the background fields) are constants that cancel in the normalization of the path integral so we can ignore them. Terms that are first order in the perturbation vanish since the background fields are assumed to satisfy the equations of motion. Therefore we need only terms that are quadratic or higher in the perturbation. 

We want to extract from this expression an on-shell 1-to-1 scattering amplitude using the standard LSZ formula
\begin{align}
    &i (2\pi)\delta\left(E_{\mathbf{p}}-E_{\mathbf{p}-\mathbf{q}}\right) \mathcal{M}_{\chi\rightarrow \chi}^{\text{(Full)}}\left(\mathbf{p},\mathbf{q}\right) \\
    &= (-i)^2[p_1^2-\mathfrak{m}^2+i0][p_2^2-\mathfrak{m}^2+i0]\int\text{d}^3 x_1 \text{d}^3 x_2 e^{-i\left(p_1\cdot x_1+p_2\cdot x_2\right)}\langle \chi^*(x_1)\chi(x_2)\rangle_{\overline{\sigma},\overline{A}_\mu} \biggr\vert_{\substack{p_1 \rightarrow(E_{\mathbf{p}+\mathbf{q}},\mathbf{p}+\mathbf{q})\\p_2 \rightarrow(-E_{\mathbf{p}},-\mathbf{p})\hspace{3mm}}}\nonumber.
\end{align}
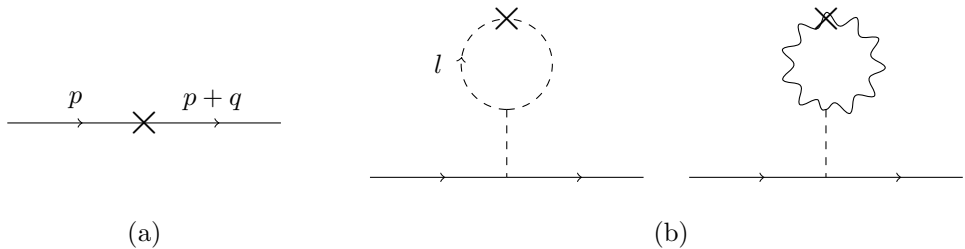
\begin{figure}
\centering
    \begin{subfigure}[b]{0.3\textwidth}
    \centering
    \begin{tikzpicture}
    \begin{scope}[scale=0.6]
    \draw[fermion] (0,-3)--(3,-3);
    \draw[fermion] (3,-3)--(6,-3);
    \node at (3,-3) {\LARGE$\times$};
    \node at (0,-4) {};
    \node at (1.5,-2.5) {\small $p$};
    \node at (4.5,-2.5) {\small $p+q$};
    \end{scope}
    \end{tikzpicture}
    \vspace{0.3cm}
    \caption{}
    \label{fig:twopointdiagramtree}
\end{subfigure}
\begin{subfigure}[b]{0.6\textwidth}
\centering
    \begin{tikzpicture}
    \begin{scope}[scale=0.6]
    \draw[fermion] (0,-3)--(3,-3);
    \draw[fermion] (3,-3)--(6,-3);
    \draw[dashed] (3,-1.5)--(3,-3);
    \draw[dashed] (3,-1.5) arc (-90:90:1);
    \draw[scalar] (3,-1.5) arc (270:90:1);
    \node at (1.5,-0.5) {\small $l$};
    \node at (3,0.5) {\LARGE$\times$};
    \end{scope}
    \begin{scope} [scale=0.6, xshift=7cm]
    \draw[fermion] (0,-3)--(3,-3);
    \draw[fermion] (3,-3)--(6,-3);
    \draw[dashed] (3,-1.5)--(3,-3);
    \draw[vector] (3,-1.5) arc (-90:270:1);
    \node at (3,0.5) {\LARGE$\times$};
    \end{scope}
\end{tikzpicture}
\vspace{0.3cm}
\caption{}
\label{fig:twopointdiagramloop}
\end{subfigure}
\caption{Diagram topologies for probe two-point function $\langle \chi^* \chi\rangle $ with $\overline{\sigma}$ and $\overline{A}_\mu$ background insertions (denoted with $\times$) at: (a) tree-level and (b) one-loop. The one-loop diagrams give a vanishing contribution in the potential region.}
    \label{fig:twopointdiagrams}
\end{figure}
At leading order, $\mathcal{O}\left(N^{1/2}\right)$, there is only a single diagram topology given in Figure \ref{fig:twopointdiagramtree}, corresponding to an insertion of the leading-order perturbative solution (\ref{pertvortexsol}). At next-to-leading order, $\mathcal{O}\left(N^{3/2}\right)$, there are in principle two different types of contribution; again from the tree-level diagram depicted in Figure \ref{fig:twopointdiagramtree} with an insertion of the $\mathcal{O}\left(N^{3/2}\right)$ perturbative solution, but also from the loop-level diagrams depicted in Figure \ref{fig:twopointdiagramloop} with insertions of the $\mathcal{O}\left(N^{1/2}\right)$ solution. We can avoid calculating such diagrams by expanding the corresponding loop integrals (in both the full-theory and REFT calculations) in the \textit{classical potential region} \cite{Bern:2019crd}. 

To define this region we need to specify our choice of scattering kinematics; we will work in the frame in which the soliton is at rest and therefore the momentum transfer is purely spatial $q^\mu = \left(0,\mathbf{q}\right)$. Denoting the incoming momentum as $p^\mu = \left(E_{\mathbf{p}},\mathbf{p}\right)$ where $E_{\mathbf{p}}\equiv \sqrt{\mathbf{p}^2+\mathfrak{m}^2}$, conservation of energy between the initial and final state implies the non-linear constraint $E_{\mathbf{p}}=E_{\mathbf{p}+\mathbf{q}}$. The loop momentum, routed as indicated in Figure \ref{fig:twopointdiagramloop}, in this non-relativistic notation is denoted $l^\mu = \left(\omega,\mathbf{l}\right)$. In this calculation we are matching the classical soliton solution to the REFT, so we expand to leading-order in the hierarchy 
\begin{equation}
    \label{probeclassical}
    M \sim \mathbf{p} \sim \mathfrak{m} \gg \mathbf{q} \sim m.
\end{equation}
Here we are using the fact that the spatial momentum transfer $\mathbf{q}$ is Fourier conjugate to the spatial separation $\mathbf{x}$ between the vortex and the probe and performing an EFT matching calculation at the scale $\mathbf{x}\sim R_{\text{interaction}} \sim m^{-1}$. Next we expand in the non-relativistic limit for the probe 
\begin{equation}
    \label{probenonrel}
    \mathfrak{m} \gg \mathbf{p},
\end{equation}
for this matching calculation it is sufficient to truncate at leading or static order. In principle we could calculate the relevant loop integrals exactly and then subsequently make the above expansions, but at high loop order this may not be feasible. Instead we make use of the method of regions and expand the loop integrals before integration \cite{Beneke:1997zp}. Moreover, we can match the contributions from a single conveniently chosen region in both full theory and the REFT instead of matching the complete result. In this case we choose to match the contributions in the classical potential region defined by the loop momentum scaling
\begin{equation}
    \label{probepotregion}
    \left(\omega,\mathbf{l}\right) \sim \left(|\mathbf{p}||\mathbf{q}|,|\mathbf{q}|\right). 
\end{equation}
In non-relativistic notation the loop integrals appearing in Figure \ref{fig:twopointdiagramloop} take the form
\begin{equation}
    \int \frac{\text{d}^{d-1}\mathbf{l}}{(2\pi)^{d-1}} \frac{\text{d}\omega}{2\pi} \frac{\mathcal{N}\left(\omega,\mathbf{l}\right)}{[\omega^2-\mathbf{l}^2-m^2+i0][\omega^2-(\mathbf{l}+\mathbf{q})^2-m^2+i0]},
\end{equation}
expanding the integrand as (\ref{probepotregion}) we find that order-by-order the resulting $\omega$-integral is scaleless and therefore vanishes in dimensional regularization. We therefore conclude that at one-loop only diagrams of the form depicted in Figure \ref{fig:twopointdiagramtree} contribute in the potential region. It is straightforward to extend this argument to higher loop order: for a loop integral to have a non-vanishing contribution in the potential region it is necessary that it should contain at least one probe propagator, but since we are explicitly working at $\mathcal{O}\left(\mathfrak{g}_{s,m}\right)$ this can never occur. 

It is therefore possible to calculate the potential region contribution to the probe amplitude in full theory at all orders in $N$, the result takes a simple form 
\begin{equation}
    \label{MprobeFullexplicit}
    \mathcal{M}_{\chi\rightarrow \chi}^{\text{(Full)}}\left(\mathbf{p},\mathbf{q}\right)\biggr\vert_{\text{potential region}} = \mathfrak{g}_s \overline{\sigma}(\mathbf{q}) - 4i \mathfrak{g}_m \sqrt{\mathbf{p}^2+\mathfrak{m}^2} \epsilon_{ij}q_i \overline{A}_j(\mathbf{q}) +\mathcal{O}\left(\mathfrak{g}_{s,m}^2\right).
\end{equation}
We now want to match this result with an equivalent calculation in the REFT. We calculate a 2-to-2 scattering amplitude between a probe $\chi$ and point-particle vortex $\Phi$ with the momentum of the incoming vortex is chosen to be at rest $P^\mu = \left(M,0\right)$ and the momentum of the probe identical to the above. Again we expand the result to leading order in (\ref{probeclassical}) and (\ref{probenonrel}), at loop-level keeping only contributions from the potential region (\ref{probepotregion}). There are two tree-level Feynman diagrams that contribute to the REFT amplitude at $\mathcal{O}\left(N^{1/2}\right)$ depicted in Figure \ref{fig:REFTprobetree}. By explicit calculation we find 
\begin{equation}
    \mathcal{M}_{\chi\Phi\rightarrow \chi\Phi}^{(\text{REFT})}\left(\mathbf{p},\mathbf{q}\right)\biggr\vert_{N^{1/2}} = \frac{\mathfrak{g}_s g_s M^{3/2}N^{1/2}}{\mathbf{q}^2+m^2}-\frac{16\mathfrak{g}_m g_m M^{3/2}N^{1/2} m\sqrt{\mathbf{p}^2+\mathfrak{m}^2}}{\mathbf{q}^2+m^2}+\mathcal{O}\left(\mathfrak{g}_{s,m}^2\right).
\end{equation}
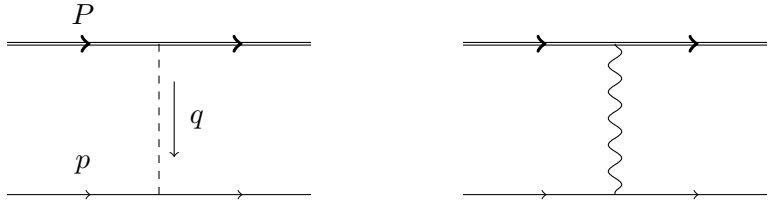
\begin{figure}
    \centering
    \begin{center}
    \begin{tikzpicture}
        \begin{scope}
        \draw[fermion,double] (0,0)--(2,0);
        \draw[fermion,double] (2,0)--(4,0);
        \draw[-,dashed] (2,0)--(2,-2);
        \draw[<-] (2.2,-1.5)--(2.2,-0.5);
        \node at (1,0.4) {$P$};
        \node at (1,-1.6) {$p$};
        \node at (2.5,-1) {$q$};
        \draw[fermion] (0,-2)--(2,-2);
        \draw[fermion] (2,-2)--(4,-2);
        \end{scope}
        \begin{scope}[xshift=6cm]
        \draw[fermion,double] (0,0)--(2,0);
        \draw[fermion,double] (2,0)--(4,0);
        \draw[vector] (2,0)--(2,-2);
        \draw[fermion] (0,-2)--(2,-2);
        \draw[fermion] (2,-2)--(4,-2);
        \end{scope}
    \end{tikzpicture}
\end{center}
    \caption{Feynman diagrams contributing to the vortex-probe scattering amplitude at $\mathcal{O}\left(N^{1/2}\right)$, the double-line denotes the vortex $\Phi$ and the single line the probe $\chi$.}
    \label{fig:REFTprobetree}
\end{figure}
Since the amplitudes we are matching between full theory and the REFT have different numbers of external particles we need to account for a (dimensionful) relative normalization factor; the explicit matching condition we are imposing is then
\begin{equation}
    \label{probematching}
    \mathcal{M}_{\chi\Phi\rightarrow \chi\Phi}^{(\text{REFT})}\left(\mathbf{p},\mathbf{q}\right)\biggr\vert_{\text{potential region}} = 2M\times \mathcal{M}_{\chi\rightarrow \chi}^{(\text{Full})}\left(\mathbf{p},\mathbf{q}\right)\biggr\vert_{\text{potential region}} +\mathcal{O}\left(\mathfrak{g}_{s,m}^2\right).
\end{equation}
To calculate the right-hand-side we need the Fourier transform of the leading-order contribution to the perturbative solution of the BPS equation (\ref{pertvortexsol}), this is found to be 
\begin{equation}
    \overline{\sigma}\left(\mathbf{q}\right) = -\frac{2\sqrt{2} M^{1/2}N^{1/2}}{\mathbf{q}^2+m^2}+\mathcal{O}\left(N^{3/2}\right), \hspace{5mm} \overline{A}_i(\mathbf{q}) = \frac{2i\sqrt{2\pi}M^{1/2}N^{1/2}m^{-1}\epsilon_{ij}q_j}{\mathbf{q}^2+m^2}+\mathcal{O}\left(N^{3/2}\right).
\end{equation}
Putting this together we find that for the matching condition to be satisfied at $\mathcal{O}\left(N^{1/2}\right)$ fixes the cubic Wilson coefficients in the REFT to take the values
\begin{equation}
    \label{probematchingresult}
    g_s = -4\sqrt{2\pi}, \hspace{10mm} g_m = -\sqrt{2\pi}.
\end{equation}
As a non-trivial consistency check of this result we note this satisfies the proportionality relation (\ref{susypredcubic}) required by the hidden $\mathcal{N}=2$ supersymmetry.

\subsection{Classical Solitons from the S-Matrix}
\label{sec:probeloop}

Above we used the matching condition (\ref{probematching}) at leading-order in $N$ (tree-level) to fix the cubic interactions of the REFT, but this relation should be true also at higher loop orders. Naively we might think that matching at higher-order in $N$ will fix further Wilson coefficients in the REFT action, in particular those corresponding to quartic and higher interactions, but this is not the case. By an argument identical to the one given above, any probe-vortex diagram in the REFT containing quartic interactions between the vortex and the mediators, and also $\mathcal{O}\left(\mathfrak{g}_{s,m}\right)$, will give a vanishing contribution in the potential region. The problem of matching higher-loop terms in (\ref{probematching}) is therefore over-constrained and provides a highly non-trivial consistency check on the REFT construction. Phrased a different way, the cubic vortex-mediator interactions determined above, together with the self-interactions of the mediator fields in the critical AHM, completely determine the (perturbative) static soliton solution (\ref{pertvortexsol}). 

The purpose of this short section is to explain how to calculate the higher-order corrections to the solution directly from the Feynman diagram expansion of the REFT probe amplitudes. This calculation is similar in spirit to the well-known method for perturbatively reconstructing the non-linear Schwarzschild solution from Feynman diagrams or directly from gauge invariant scattering amplitudes \cite{Duff:1973zz,Neill:2013wsa,Bjerrum-Bohr:2018xdl,Jakobsen:2020ksu,Mougiakakos:2020laz}.

To see how this works in practice we begin by inverting the explicit expression for the full theory probe amplitude (\ref{MprobeFullexplicit}) to derive a simple expression for the classical solution in terms of the REFT probe amplitude
\begin{align}
    \label{classicalsolfromamp}
    \overline{\sigma}(\mathbf{x}) &= \frac{1}{2M}\int\frac{\text{d}^{2}\mathbf{q}}{(2\pi)^{2}} e^{i\mathbf{q}\cdot\mathbf{x}}\mathcal{M}_{\chi\Phi\rightarrow \chi\Phi}^{(\text{REFT})}\left(0,\mathbf{q}\right)\biggr\vert_{\mathfrak{g}_{s},\;\text{potential region}} \nonumber\\
    \overline{A}_i(\mathbf{x}) &= -\frac{i}{8\mathfrak{m}M}\int\frac{\text{d}^{2}\mathbf{q}}{(2\pi)^{2}} \frac{\epsilon_{ij}q_j}{\mathbf{q}^2}e^{i\mathbf{q}\cdot\mathbf{x}}\mathcal{M}_{\chi\Phi\rightarrow \chi\Phi}^{(\text{REFT})}\left(0,\mathbf{q}\right)\biggr\vert_{\mathfrak{g}_{m},\;\text{potential region}}.
\end{align}
\begin{figure}
    \centering
    \begin{tikzpicture}
    \begin{scope}[scale=0.55]
    \draw[fermion,double] (0,0)--(1.5,0);
    \draw[fermion,double] (1.5,0)--(4.5,0);
    \draw[fermion,double] (4.5,0)--(6,0);
    \draw[fermion] (0,-3)--(3,-3);
    \draw[fermion] (3,-3)--(6,-3);
    \draw[dashed] (1.5,0)--(3,-1.5);
    \draw[dashed] (4.5,0)--(3,-1.5);
    \draw[dashed] (3,-1.5)--(3,-3);
    \draw[->] (2,-1.2)--(1.4,-0.6);
    \draw[->] (3.3,-1.8)--(3.3,-2.6);
    \node at (3.8,-2.2) {$q$};
    \node at (0.75,0.7) {$P$};
    \node at (1.4,-1.3) {$l$};
    \node at (1.5,-3.8) {$p$};
    \end{scope}
    \begin{scope} [scale=0.55, xshift=7cm]
    \draw[fermion,double] (0,0)--(1.5,0);
    \draw[fermion,double] (1.5,0)--(4.5,0);
    \draw[fermion,double] (4.5,0)--(6,0);
    \draw[fermion] (0,-3)--(3,-3);
    \draw[fermion] (3,-3)--(6,-3);
    \draw[vector] (1.5,0)--(3,-1.5);
    \draw[vector] (4.5,0)--(3,-1.5);
    \draw[dashed] (3,-1.5)--(3,-3);
    \end{scope}
    \begin{scope}[scale=0.55, xshift=14cm]
    \draw[fermion,double] (0,0)--(1.5,0);
    \draw[fermion,double] (1.5,0)--(4.5,0);
    \draw[fermion,double] (4.5,0)--(6,0);
    \draw[fermion] (0,-3)--(3,-3);
    \draw[fermion] (3,-3)--(6,-3);
    \draw[vector] (1.5,0)--(3,-1.5);
    \draw[dashed] (4.5,0)--(3,-1.5);
    \draw[vector] (3,-1.5)--(3,-3);
    \end{scope}
    \begin{scope} [scale=0.55, xshift=21cm]
    \draw[fermion,double] (0,0)--(1.5,0);
    \draw[fermion,double] (1.5,0)--(4.5,0);
    \draw[fermion,double] (4.5,0)--(6,0);
    \draw[fermion] (0,-3)--(3,-3);
    \draw[fermion] (3,-3)--(6,-3);
    \draw[dashed] (1.5,0)--(3,-1.5);
    \draw[vector] (4.5,0)--(3,-1.5);
    \draw[vector] (3,-1.5)--(3,-3);
    \end{scope}
\end{tikzpicture}
    \caption{Feynman diagrams contributing to the vortex-probe amplitude at $\mathcal{O}\left(N^{3/2}\right)$.}
    \label{fig:REFTprobeloop}
\end{figure}
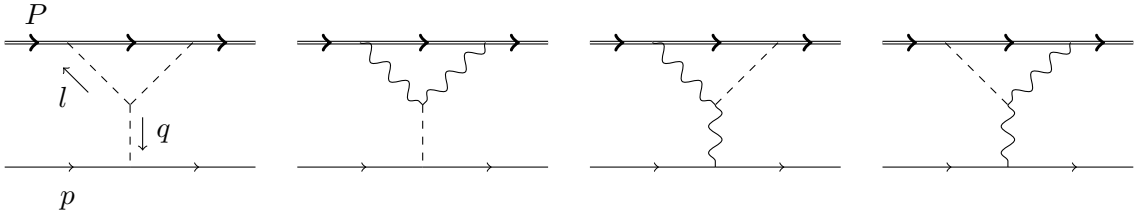
At $L$-loops the only diagrams that give a non-vanishing contribution are the so-called ``fan diagrams" containing a single cubic probe coupling and $L+1$ cubic interactions between the vortex and the mediators. At one-loop or $\mathcal{O}\left(N^{3/2}\right)$ there are four such diagrams depicted in Figure \ref{fig:REFTprobeloop}. Expanding these diagrams in the classical potential region we find
\begin{align}
\label{oneloopvortexprobe}
&\mathcal{M}_{\chi\Phi\rightarrow \chi\Phi}^{(\text{REFT})}\left(0,\mathbf{q}\right)\bigg\vert_{N^{3/2},\;\text{potential region}}  \nonumber\\
&= 4\sqrt{2}\pi^{3/2}N^{3/2} M^{3/2}\left(\mathfrak{g}_s  -8\mathfrak{g}_m m \mathfrak{m}\right) \int \frac{\text{d}^{2}\mathbf{l}}{(2\pi)^2} \frac{1}{[\mathbf{l}^2+m^2][\left(\mathbf{l}+\mathbf{q}\right)^2+m^2]} \\
&\hspace{5mm}-8\sqrt{2}\pi^{3/2} N^{3/2} m^2 M^{3/2}\left(\mathfrak{g}_s -4\mathfrak{g}_m m \mathfrak{m}\right) \frac{1}{\mathbf{q}^2+m^2}\int \frac{\text{d}^{2}\mathbf{l}}{(2\pi)^2} \frac{1}{[\mathbf{l}^2+m^2][\left(\mathbf{l}+\mathbf{q}\right)^2+m^2]}\nonumber.
\end{align}
More information about the general strategy for simplification of these expression is given in Section \ref{sec:loop}. Inserting these expressions into (\ref{classicalsolfromamp}) and making use of Fourier transforms derived in Appendix \ref{sec:Fourier} we find that the result exactly matches the $\mathcal{O}\left(N^{3/2}\right)$ perturbative solution of the BPS equations (\ref{pertvortexsol}). 

\section{Non-Relativistic Effective Field Theory}
\label{sec:NREFTsec}

\subsection{Definition of the NREFT}
\label{sec:NREFT}

In this paper we will focus on conservative, two-body interactions between identical vortices with winding number $N$. This means we focus on quartic effective interactions generated by integrating out mediator potential modes. Higher-multiplicity interactions, as well as dissipative effects (importantly including radiation) correspond to further effective interactions, the calculation of which we leave to future work.

\begin{figure}
    \centering
    \begin{tikzpicture}
    \begin{scope}
    \draw[fermion,double] (-1,1)--(0,0);
    \draw[fermion,double] (-1,-1)--(0,0);
    \draw[fermion,double] (0,0)--(1,1);
    \draw[fermion,double] (0,0)--(1,-1);
    \node at (1.5,0) {+};
    \node at (-0.5,0.95) {$\mathbf{p}$};
    \node at (-0.5,-1.05) {$-\mathbf{p}$};
    \node at (0.5,1) {$\mathbf{p}'$};
    \node at (0.4,-1) {$-\mathbf{p}'$};
    \end{scope}
    \begin{scope}[xshift=3cm]
    \draw[fermion,double] (-1,1)--(0,0);
    \draw[fermion,double] (-1,-1)--(0,0);
    \draw[fermion,double] (0,0) arc (150:30:1);
    \draw[fermion,double] (0,0) arc (-150:-30:1);
    \draw[fermion,double] (1.732,0)--(2.732,1);
    \draw[fermion,double] (1.732,0)--(2.732,-1);
    \node at (3.25,0) {+};
    \node at (0.9,1) {$\mathbf{k}$};
    \node at (0.8,-1) {$-\mathbf{k}$};
    \end{scope}
    \begin{scope}[xshift=7.7cm]
    \draw[fermion,double] (-1,1)--(0,0);
    \draw[fermion,double] (-1,-1)--(0,0);
    \draw[fermion,double] (0,0) arc (150:30:1);
    \draw[fermion,double] (0,0) arc (-150:-30:1);
    \draw[fermion,double] (1.732,0) arc (150:30:1);
    \draw[fermion,double] (1.732,0) arc (-150:-30:1);
    \draw[fermion,double] (3.464,0)--(4.464,1);
    \draw[fermion,double] (3.464,0)--(4.464,-1);
    \node at (5,0) {+};
    \node at (6,0) {$\cdots$};
    \end{scope}
\end{tikzpicture}
    \caption{Feynman diagrams contributing to the conservative two-body vortex-vortex scattering amplitude in the NREFT.}
    \label{fig:NREFTbubble}
\end{figure}
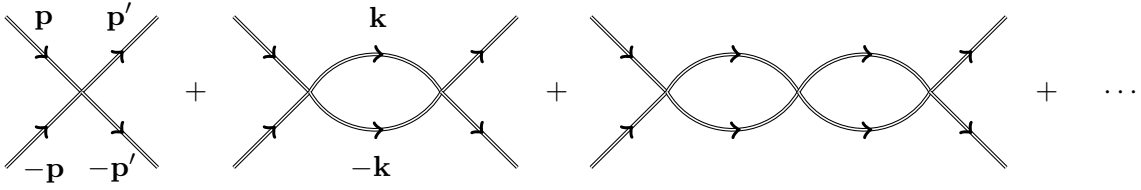

The conservative two-body sector of the NREFT is defined by an effective action of the form \cite{Neill:2013wsa,Cheung:2018wkq}
\begin{align}  
\label{SNREFT}
    S_{\text{NREFT}}\left[\Phi\right] &= \int \text{d}t \left[\int\frac{\text{d}^2\mathbf{k}}{(2\pi)^2} \Phi^\dagger(-\mathbf{k})\left(i\partial_t-\sqrt{\mathbf{k}^2+M^2}\right)\Phi(\mathbf{k}) \right. \nonumber\\
    &\hspace{15mm}\left.-\int\frac{\text{d}^2\mathbf{k}}{(2\pi)^2}\int\frac{\text{d}^2\mathbf{k}'}{(2\pi)^2}V\left(\mathbf{k},\mathbf{k}'\right)\Phi^\dagger(\mathbf{k}')\Phi(\mathbf{k})\Phi^\dagger(-\mathbf{k}')\Phi(-\mathbf{k})\right],
\end{align}
where $V(\mathbf{k},\mathbf{k}')$ corresponds to the effective vortex-vortex potential.\footnote{Throughout this section we will abuse notation and use the same symbol to denote the potential as a function of the ``incoming" and ``outgoing" momenta $V\left(\mathbf{p},\mathbf{p}'\right)=V\left(\mathbf{p},\mathbf{p}-\mathbf{q}\right)$ as well as a function of instantaneous momentum and position $V\left(\mathbf{p},\mathbf{x}\right)$, where $\mathbf{q}$ is the Fourier conjugate variable to $\mathbf{x}$; it should be clear from context which is intended.} Scattering amplitudes in the NREFT are calculated using Feynman rules derived from the effective action (\ref{SNREFT}), as given in Appendix \ref{sec:Feynman}. In particular, for conservative two-body scattering the only diagrams that contribute are iterated bubble diagrams depicted in Figure \ref{fig:NREFTbubble}.
In this section the NREFT amplitude is calculated in the center-of-mass frame defined by the kinematics
\begin{equation}
\label{comkin}
    p_1 = \left(E_{\mathbf{p}},\mathbf{p}\right), \hspace{5mm} p_2 = \left(E_{\mathbf{p}},-\mathbf{p}\right), \hspace{5mm} p_1' = \left(E_{\mathbf{p}},\mathbf{p}'\right), \hspace{5mm} p_2' = \left(E_{\mathbf{p}},-\mathbf{p}'\right),
\end{equation}
where $\mathbf{p}'=\mathbf{p}-\mathbf{q}$. After calculating the trivial $\omega$-integrals, the $L$-loop contribution to the NREFT amplitude is given by
\begin{equation}
    M_{\text{NREFT}}^{L\text{-loop}}\left(\mathbf{p},\mathbf{p}'\right) = -\int_{\mathbf{k}_1...\mathbf{k}_L} V\left(\mathbf{p},\mathbf{k}_1\right)\Delta_{\Phi\Phi}\left(\mathbf{p},\mathbf{k}_1\right)\cdots\Delta_{\Phi\Phi}\left(\mathbf{k}_{L-1},\mathbf{k}_L\right)V\left(\mathbf{k}_L,\mathbf{p}'\right),
\end{equation}
where 
\begin{equation}
    \Delta_{\Phi\Phi}\left(\mathbf{l},\mathbf{l}'\right) = \frac{1}{2E_{\mathbf{l}}-2E_{\mathbf{l}'}+i0},\hspace{10mm} \int_{\mathbf{k}} = \int \frac{\text{d}^{d-1}\mathbf{k}}{(2\pi)^{d-1}}.
\end{equation}
Formally this is the same as the Born series solution of the relativistic Lippmann-Schwinger equation \cite{Lippmann:1950zz,Cristofoli:2019neg}
\begin{equation}
    \label{Lippmann}
    M_{\text{NREFT}}\left(\mathbf{p},\mathbf{p}'\right) = -V\left(\mathbf{p},\mathbf{p}'\right)+\int_{\mathbf{k}} \frac{V\left(\mathbf{p},\mathbf{k}\right)M_{\text{NREFT}}\left(\mathbf{k},\mathbf{p}'\right)}{2E_{\mathbf{p}}-2E_{\mathbf{k}}+i0},
\end{equation}
derived from an effective Hamiltonian
\begin{equation}
    \label{HNREFT}
    H_{\text{NREFT}}\left(\mathbf{p}_1,\mathbf{p}_2,\mathbf{x}_1,\mathbf{x}_2\right) = \sum_{i=1}^2\sqrt{\mathbf{p}_i^2+M^2} + V\left(\mathbf{p}_1,\mathbf{p}_2,\mathbf{x}_1,\mathbf{x}_2\right).
\end{equation}
In this formalism the potential is a Wilson coefficient that is calculated by matching with the non-relativistic expansion of the REFT. By standard convention, the amplitudes calculated in the (N)REFT are (non-)relativistically normalized and so the matching condition takes the form
\begin{equation}
    \label{matchingNREFT}
    M_{\text{NREFT}}\left(\mathbf{p},\mathbf{p}'\right) = \frac{\mathcal{M}_{\text{REFT}}\left(\mathbf{p},\mathbf{p}'\right)}{4\left(\mathbf{p}^2+M^2\right)}.
\end{equation}
This matching is performed order-by-order in $N$; in the NREFT this means the potential has an expansion of the form
\begin{equation}
     V\left(\mathbf{p},\mathbf{p}'\right) = \sum_{i=1}^\infty V^{(i)}\left(\mathbf{p},\mathbf{p}'\right),
\end{equation}
where $V^{(i)}\sim N^i$. At leading-order, or $\mathcal{O}\left(N\right)$, the potential is given straightforwardly by the first Born approximation 
\begin{equation}
    \label{matching1}
    V^{(1)}\left(\mathbf{p},\mathbf{p}'\right) = -\frac{\mathcal{M}^{(1)}_{\text{REFT}}\left(\mathbf{p},\mathbf{p}'\right)}{4\left(\mathbf{p}^2+M^2\right)}.
\end{equation}
At next-to-leading-order, or $\mathcal{O}\left(N^2\right)$, the potential is given by rearranging the matching condition (\ref{matchingNREFT}) 
\begin{equation}
    \label{matching2}
    V^{(2)}\left(\mathbf{p},\mathbf{p}'\right) = -\frac{\mathcal{M}^{(2)}_{\text{REFT}}\left(\mathbf{p},\mathbf{p}'\right)}{4\left(\mathbf{p}^2+M^2\right)} - \int \frac{\text{d}^{d-1}\mathbf{k}}{(2\pi)^{d-1}}\frac{V^{(1)}\left(\mathbf{p},\mathbf{k}\right)V^{(1)}\left(\mathbf{k},\mathbf{p}'\right)}{2E_{\mathbf{p}}-2E_{\mathbf{k}}+i0},
\end{equation}
the second term is variously called an \textit{iteration} contribution or a \textit{Born subtraction}.

As in the probe calculation above, the matching is performed by expanding loop integrals in the potential region. In practice it is convenient to expand the REFT amplitudes in two steps following an approach introduced in \cite{Parra-Martinez:2020dzs}. In the first step we make a relativistic expansion in the \textit{soft region} defined by 
\begin{equation}
	\label{soft}
    M\sim p \gg l \sim q \sim m,
\end{equation}
where we perform the calculation in a general reference frame. As we will see this allows us to make use of standard Lorentz covariant integral reduction techniques to simplify expressions in terms of a set of scalar soft integrals with linearized vortex propagators. The second step is to then fix center-of-mass frame kinematics (\ref{comkin}) and expand the soft integrals non-relativistically in the potential region
\begin{equation}
    \label{potreg}
    \left(\omega,\mathbf{l}\right) \sim \left(|\mathbf{p}||\mathbf{q}|,|\mathbf{p}|\right).
\end{equation}
As described in detail in Appendix \ref{sec:soft}, in all cases considered in this paper the resulting velocity series can be resummed, giving the complete velocity dependence of the vortex-vortex potential up to $\mathcal{O}\left(N^2\right)$. For the iteration contributions to the NREFT amplitudes we align the loop integration by defining $\mathbf{k}=\mathbf{l}+\mathbf{p}$, and then expand in the classical potential region (\ref{probepotregion}).

Finally, there is an ambiguity in the potential defined using (\ref{Lippmann}) arising from the need to continue the amplitude away from the implicit constraint surface of energy conservation \cite{Jones:2022aji}. We need to choose a resolution of the non-linear constraint $E_{\mathbf{p}} = E_{\mathbf{p}-\mathbf{q}}$, different choices lead to formally distinct classical potentials. In the Hamiltonian formalism (\ref{HNREFT}) these different choices are related by some canonical transformation on the phase space \cite{Bern:2019crd}, while in the field theoretic NREFT formalism (\ref{SNREFT}) they correspond to field redefinitions. The different choices of potential therefore lead to identical sets of physical observables and so the choice is one of convenience. In this case there is a canonical choice that leads to very simple expressions: in the center-of-mass frame, the energy conservation constraint is equivalent to the condition
\begin{equation}
    \mathbf{p}\cdot \mathbf{q} = \frac{1}{2}\mathbf{q}^2,
\end{equation}
which we can use to rewrite all appearances of $\mathbf{p}\cdot \mathbf{q}$ in the amplitudes before matching. All potentials calculated in this paper are defined in this way, a choice that in the post-Minkowskian literature is usually called \textit{isotropic gauge} \cite{Cheung:2018wkq,Bjerrum-Bohr:2018xdl,Bern:2019crd}. 

\subsection{Tree-Level Potential}
\label{sec:tree}

We begin with the calculation of the vortex-vortex potential at tree-level or $\mathcal{O}\left(N\right)$. In the REFT there are two Feynman diagrams that contribute, corresponding to the exchange of the Higgs boson $\sigma$ and the massive photon $\gamma$ as depicted in Figure \ref{fig:treeclassical}. It is instructive to work through this simple example in some detail as a warm-up for the more complicated one-loop calculation in the following subsection. Using the Feynman rules given in Appendix \ref{sec:Feynman} and the result of the probe matching calculation (\ref{probematchingresult}) the REFT tree amplitude can be written in the form
\begin{figure}
\centering
\begin{subfigure}[b]{0.6\textwidth}
    \centering
    \begin{tikzpicture}
        \begin{scope}
        \draw[fermion,double] (0,0)--(2,0);
        \draw[fermion,double] (2,0)--(4,0);
        \draw[-,dashed] (2,0)--(2,-2);
        \draw[<-] (1.8,-1.5)--(1.8,-0.5);
        \node at (1,0.6) {$p_1$};
        \node at (1,-2.6) {$p_2$};
        \node at (1.6,-1) {$q$};
        \node at (2.4,-1) {$\sigma$};
        \draw[fermion,double] (0,-2)--(2,-2);
        \draw[fermion,double] (2,-2)--(4,-2);
        \end{scope}
        \begin{scope}[xshift=5cm]
        \draw[fermion,double] (0,0)--(2,0);
        \draw[fermion,double] (2,0)--(4,0);
        \draw[vector] (2,0)--(2,-2);
        \node at (2.4,-1) {$\gamma$};
        \draw[fermion,double] (0,-2)--(2,-2);
        \draw[fermion,double] (2,-2)--(4,-2);
        \end{scope}
    \end{tikzpicture}
    \caption{}
    \label{fig:treeclassical}
\end{subfigure}
\begin{subfigure}[b]{0.35\textwidth}
    \centering
    \begin{tikzpicture}
        \begin{scope}
        \draw[fermion,double] (0,0)--(2,0);
        \draw[fermion,double] (2,0)--(4,0);
        \draw[-,dashed] (2,0)--(2,-2);
        \node at (1,0.6) {};
        \node at (1,-2.75) {};
        \node at (2.4,-1) {$\tilde{\sigma}$};
        \draw[fermion,double] (0,-2)--(2,-2);
        \draw[fermion,double] (2,-2)--(4,-2);
        \end{scope}
    \end{tikzpicture}
    \caption{}
    \label{fig:treequantum}
\end{subfigure}
    \caption{Feynman diagrams contributing to $\mathcal{O}\left(N\right)$ vortex-vortex potential. Only diagrams (a) corresponding to Higgs and photon exchange are necessary in the classical limit. In the fully quantum mechanical calculation we must also include diagram (b) corresponding to exchange of the additional pseudoscalar $\tilde{\sigma}$ required by $\mathcal{N}=2$ supersymmetry to produce a cancellation of static forces.}
\end{figure}
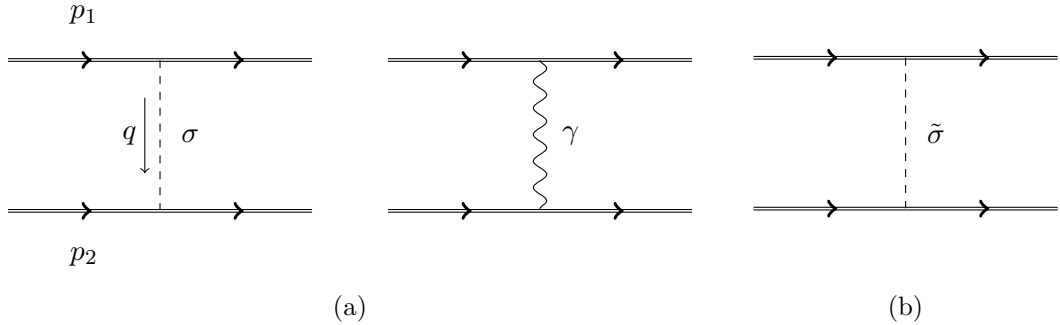
\begin{equation}
    \label{M1REFTexpanded}
    \mathcal{M}^{(1)} = -\frac{32\pi MN \left(M^2-(p_1\cdot p_2)\right)}{q^2-m^2}+\frac{8\pi MN m^2}{q^2-m^2}+\frac{32\pi MN}{m^2}\left(p_1\cdot p_2 + \frac{1}{4}\left(q^2+m^2\right)\right).
\end{equation}
As we will see, the first term corresponds to the long-range classical potential, the numerator factor vanishes in the static limit as we expect for BPS vortices. The second term also generates a long-range potential, but it is suppressed relative to the first term by a factor $\left(\frac{m}{M}\right)^2$ and therefore does not contribute to the classical potential. The third term corresponds to a contact or short-range interaction, in the corresponding position space potential it generates terms proportional to $\delta^{(2)}\left(\mathbf{x}\right)$ and $\nabla^2\delta^{(2)}\left(\mathbf{x}\right)$; from now on we will drop such terms when they appear. Using (\ref{matching1}) the classical long-range potential at $\mathcal{O}\left(N\right)$ is given in momentum space as
\begin{equation}
	\label{Vtree}
    V^{(1)}\left(\mathbf{p},\mathbf{p}'\right) = \frac{16\pi MN \mathbf{p}^2}{\left[\mathbf{p}^2+M^2\right]\left[|\mathbf{p}-\mathbf{p}'|^2+m^2\right]},
\end{equation}
and in position space
\begin{equation}
    V^{(1)}\left(\mathbf{p},\mathbf{x}\right) = \frac{8MN \mathbf{p}^2}{\mathbf{p}^2+M^2}K_0\left(mr\right).
\end{equation}
Before proceeding to one-loop, it is interesting to consider extending this classical calculation to the full quantum long-range potential; this means keeping the second term in (\ref{M1REFTexpanded}). Here we find a puzzle, this term clearly does not vanish in the static limit and so seems to violate our expectations about forces between critical vortices. The resolution is to be found in the analysis of Section \ref{sec:susy}: the cancellation of static forces for quantum vortices follows from their identification with BPS states in the $\mathcal{N}=2$ supersymmetric AHM. But as shown in (\ref{cubicvortexscalarint}) the supersymmetric model contains an additional pseudoscalar boson $\tilde{\sigma}$ that only decouples in the strict classical limit. In the quantum mechanical and supersymmetric version of this calculation we should also include this degree-of-freedom and therefore the additional exchange diagram depicted in Figure \ref{fig:treequantum}. Explicitly this generates an additional contribution to the REFT amplitude
\begin{equation}
    \mathcal{M}^{(1)}\biggr\vert_{\mathcal{N}=2\; \text{AHM}} = \mathcal{M}^{(1)}\biggr\vert_{\text{critical AHM}} -\frac{8\pi MN m^2}{q^2-m^2},
\end{equation}
which precisely cancels the second term in (\ref{M1REFTexpanded}) and therefore restores the expected cancellation of static forces. As a consequence, the quantum long-range potential is identical to the classical long-range potential at this order.

\subsection{One-Loop Potential}
\label{sec:loop}

In this section we present the calculation of the one-loop or $\mathcal{O}\left(N^2\right)$ classical vortex-vortex potential. Below we will calculate the \textit{triangle}, \textit{(crossed-)box} and \textit{iteration} pieces individually and assemble them at the end.\footnote{Various one-loop diagram toplogies are not described in this section, such as \textit{bubble} diagrams and \textit{mushroom} diagrams. We have explicitly checked that such diagrams give a vanishing contribution in the classical potential region.}

\subsubsection*{Triangle Diagrams}

\begin{figure}
    \centering
    \begin{tikzpicture}
    \begin{scope}[scale=0.55]
    \draw[fermion,double] (0,0)--(1.5,0);
    \draw[fermion,double] (1.5,0)--(4.5,0);
    \draw[fermion,double] (4.5,0)--(6,0);
    \draw[fermion,double] (0,-3)--(3,-3);
    \draw[fermion,double] (3,-3)--(6,-3);
    \draw[dashed] (1.5,0)--(3,-1.5);
    \draw[dashed] (4.5,0)--(3,-1.5);
    \draw[dashed] (3,-1.5)--(3,-3);
    \draw[->] (2,-1.2)--(1.4,-0.6);
    \draw[->] (3.3,-1.8)--(3.3,-2.6);
    \node at (3.6,-2.2) {$q$};
    \node at (0.75,0.6) {$p_1$};
    \node at (1.5,-1.2) {$l$};
    \node at (1.5,-3.6) {$p_2$};
    \end{scope}
    \begin{scope} [scale=0.55, xshift=7cm]
    \draw[fermion,double] (0,0)--(1.5,0);
    \draw[fermion,double] (1.5,0)--(4.5,0);
    \draw[fermion,double] (4.5,0)--(6,0);
    \draw[fermion,double] (0,-3)--(3,-3);
    \draw[fermion,double] (3,-3)--(6,-3);
    \draw[vector] (1.5,0)--(3,-1.5);
    \draw[vector] (4.5,0)--(3,-1.5);
    \draw[dashed] (3,-1.5)--(3,-3);
    \end{scope}
    \begin{scope} [scale=0.55, xshift=14cm]
    \draw[fermion,double] (0,0)--(1.5,0);
    \draw[fermion,double] (1.5,0)--(4.5,0);
    \draw[fermion,double] (4.5,0)--(6,0);
    \draw[fermion,double] (0,-3)--(3,-3);
    \draw[fermion,double] (3,-3)--(6,-3);
    \draw[vector] (1.5,0)--(3,-1.5);
    \draw[dashed] (4.5,0)--(3,-1.5);
    \draw[vector] (3,-1.5)--(3,-3);
    \end{scope}
    \begin{scope} [scale=0.55, xshift=21cm]
    \draw[fermion,double] (0,0)--(1.5,0);
    \draw[fermion,double] (1.5,0)--(4.5,0);
    \draw[fermion,double] (4.5,0)--(6,0);
    \draw[fermion,double] (0,-3)--(3,-3);
    \draw[fermion,double] (3,-3)--(6,-3);
    \draw[dashed] (1.5,0)--(3,-1.5);
    \draw[vector] (4.5,0)--(3,-1.5);
    \draw[vector] (3,-1.5)--(3,-3);
    \end{scope}
\end{tikzpicture}
    \caption{Triangle diagrams contributing to the $\mathcal{O}\left(N^2\right)$ vortex-vortex potential. }
    \label{fig:triangle}
\end{figure}
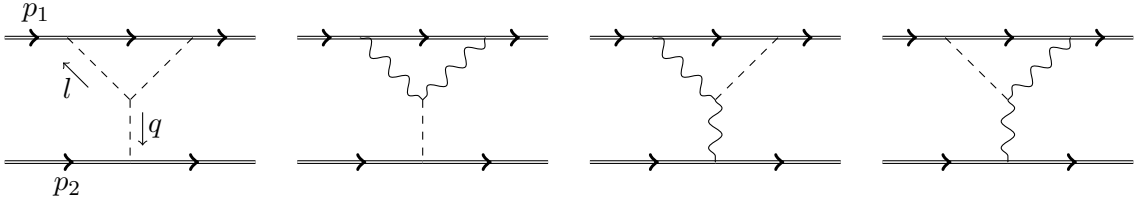
The most complicated part of the potential is generated by the triangle diagrams depicted in Figure \ref{fig:triangle}. Summing these graphs, expanding to leading-order in the soft expansion (\ref{soft}) and subsequently reducing tensor numerators, we find the result 
\begin{align}
    \label{triangleamplexpanded}
    \mathcal{M}^{(2)}_{\tri{}} &= -\frac{128i\pi^2 M^2 N^2 m^2\left(M^2-(p_1\cdot p_2)\right)}{q^2-m^2} \int \frac{\text{d}^d l}{(2\pi)^d} \frac{1}{[l^2-m^2][(l+q)^2-m^2][p_1\cdot l + i0]} \nonumber\\
    &\hspace{5mm}-64i\pi^2 M^2 N^2\left(M^2-2(p_1\cdot p_2)\right)\int \frac{\text{d}^d l}{(2\pi)^d} \frac{1}{[l^2-m^2][(l+q)^2-m^2][p_1\cdot l + i0]} \nonumber\\
    &\hspace{5mm}+\frac{128i\pi^2 M^4 N^2}{q^2-m^2}\int \frac{\text{d}^d l}{(2\pi)^d} \frac{1}{[l^2-m^2][p_1\cdot l + i0]}.
\end{align}
The first two terms are UV and IR finite and generate contributions to the physical potential. The third term is logarithmically divergent\footnote{This divergence arises solely from the second diagram in Figure \ref{fig:triangle}, corresponding to a pinch of one of the photon propagators.} and so needs to be renormalized by adding a vertex counterterm of the form $\delta \mathcal{L}^{(3)}_{\text{vortex}} \supset \delta g_s M^{3/2}N^{3/2} \sigma |\Phi|^2$. In this case there is a unique renormalization scheme consistent with both the cancellation of static forces and with matching the probe limit result (\ref{oneloopvortexprobe}) to the perturbative classical solution (\ref{pertvortexsol}), this amounts to setting such divergent scalar integrals to zero when they appear.

Next we add the $1\leftrightarrow 2$ diagrams, these have an almost identical contribution to the above but with $p_1\leftrightarrow p_2$ and $q\rightarrow -q$. Actually, since both vortices are indistinguishable these two contributions are identical in the center-of-mass frame\footnote{To see this explicitly we temporarily rewrite the integral in non-relativistic notation and make the change of variables $\mathbf{l}\rightarrow -\mathbf{l}$}, therefore if we add both orientations of triangle graph together we find 
\begin{align}
   &\mathcal{M}^{(2)}_{\tri{}}+\mathcal{M}^{(2)}_{\triinv{}} \nonumber\\
   &= -\frac{256i\pi^2 M^2 N^2 m^2\left(M^2-(p_1\cdot p_2)\right)}{q^2-m^2} \int \frac{\text{d}^d l}{(2\pi)^d} \frac{1}{[l^2-m^2][(l+q)^2-m^2][p_1\cdot l + i0]} \nonumber\\
    &\hspace{5mm}-128i\pi^2 M^2 N^2\left(M^2-2(p_1\cdot p_2)\right)\int \frac{\text{d}^d l}{(2\pi)^d} \frac{1}{[l^2-m^2][(l+q)^2-m^2][p_1\cdot l + i0]}.
\end{align}
Here we find a problem, while the first term manifestly vanishes in the static limit the second does not. As we will see in the following, the remaining diagrams do not fix this problem. Instead this is an indication that we have missed a contribution from a \textit{seagull} vertex. There are many possible contact terms, parameterizing finite size corrections to the vortex and must be calculated from a separate matching calculation. Here we use the weaker constraint that the vortex-vortex potential should vanish in the static limit.

The simplest seagull term we can add to restore the cancellation of static force has the form
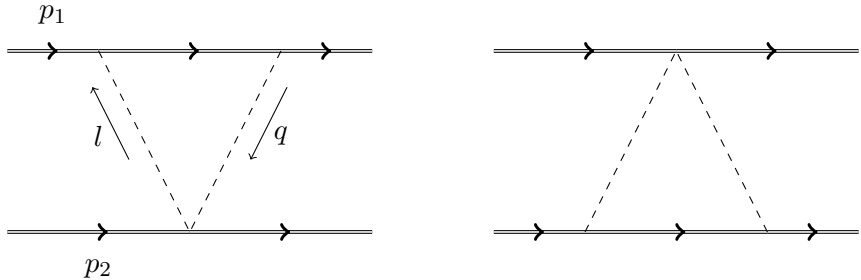
\begin{figure}
    \centering
    \begin{tikzpicture}
    \begin{scope}[scale=0.8]
    \draw[fermion,double] (0,0)--(1.5,0);
    \draw[fermion,double] (1.5,0)--(4.5,0);
    \draw[fermion,double] (4.5,0)--(6,0);
    \draw[fermion,double] (0,-3)--(3,-3);
    \draw[fermion,double] (3,-3)--(6,-3);
    \draw[dashed] (1.5,0)--(3,-3);
    \draw[dashed] (4.5,0)--(3,-3);
    \draw[->] (2,-1.8)--(1.4,-0.6);
    \draw[->] (4.6,-0.6)--(4,-1.8);
    \node at (4.5,-1.4) {$q$};
    \node at (0.75,0.6) {$p_1$};
    \node at (1.5,-1.4) {$l$};
    \node at (1.5,-3.6) {$p_2$};
    \end{scope}
    \begin{scope} [scale=0.8, xshift=8cm]
    \draw[fermion,double] (0,-3)--(1.5,-3);
    \draw[fermion,double] (1.5,-3)--(4.5,-3);
    \draw[fermion,double] (4.5,-3)--(6,-3);
    \draw[fermion,double] (0,0)--(3,0);
    \draw[fermion,double] (3,0)--(6,0);
    \draw[dashed] (1.5,-3)--(3,0);
    \draw[dashed] (4.5,-3)--(3,0);
    \end{scope}
\end{tikzpicture}
    \caption{``Seagull" diagrams contributing to the $\mathcal{O}\left(N^2\right)$ vortex-vortex potential, necessary for the cancellation of static forces. }
    \label{fig:seagull}
\end{figure}
\begin{equation}
	\label{Svortex4}
    S^{(4)}_\text{vortex}\left[\sigma,A_\mu,\Phi\right] \supset \int \text{d}^3 x \biggr[g_{\sigma\sigma} MN \sigma^2 |\Phi|^2\biggr],
\end{equation}
which contributes two additional triangle graphs depicted in Figure \ref{fig:seagull}. Explicitly we find that we need
\begin{equation}
    \label{gss}
    g_{\sigma\sigma} = -2\pi.
\end{equation}
Putting this together we find the complete triangle contribution to the soft REFT amplitude 
\begin{align}
	\label{MREFTtri}
    &\mathcal{M}^{(2)}_{\tri{}}+\mathcal{M}^{(2)}_{\triinv{}}+\mathcal{M}^{(2)}_{\tric{}}+\mathcal{M}^{(2)}_{\tricinv{}} \nonumber\\
   &= -\frac{256i\pi^2 M^2 N^2 m^2\left(M^2-(p_1\cdot p_2)\right)}{q^2-m^2} \int \frac{\text{d}^d l}{(2\pi)^d} \frac{1}{[l^2-m^2][(l+q)^2-m^2][p_1\cdot l + i0]} \nonumber\\
    &\hspace{5mm}-256i\pi^2 M^2 N^2\left(M^2-(p_1\cdot p_2)\right)\int \frac{\text{d}^d l}{(2\pi)^d} \frac{1}{[l^2-m^2][(l+q)^2-m^2][p_1\cdot l + i0]}.
\end{align}
The next step is to fix center-of-mass kinematics (\ref{comkin}) and expand the remaining integrals in the potential region (\ref{potreg}). As shown in detail in Appendix \ref{sec:soft} the resulting expansion can be resummed, giving contributions to the potential region with exact velocity dependence. The complete triangle contribution to the momentum-space potential is then 
\begin{align}
	\label{Vtri}
   &V^{(2)}_{\tri{}}+V^{(2)}_{\triinv{}}+V^{(2)}_{\tric{}}+V^{(2)}_{\tricinv{}}\nonumber\\
   &= -\frac{64\pi^2 M N^2\mathbf{p}^2}{\mathbf{p}^2+M^2}\int \frac{\text{d}^2 \mathbf{l}}{(2\pi)^2} \frac{1}{[\mathbf{l}^2+m^2][(\mathbf{l}+\mathbf{q})^2+m^2]} \nonumber\\
   &\hspace{5mm}+\frac{64\pi^2 M N^2\mathbf{p}^2}{\mathbf{p}^2+M^2}\left(\frac{1}{\mathbf{q}^2+m^2}\int \frac{\text{d}^2 \mathbf{l}}{(2\pi)^2} \frac{m^2}{[\mathbf{l}^2+m^2][(\mathbf{l}+\mathbf{q})^2+m^2]}\right). 
\end{align}

\subsubsection*{Box and Crossed-Box Diagrams}

\begin{figure}
    \centering
    \begin{tikzpicture}
    \begin{scope}[scale=0.55]
    \draw[fermion,double] (0,0)--(1.5,0);
    \draw[fermion,double] (1.5,0)--(4.5,0);
    \draw[fermion,double] (4.5,0)--(6,0);
    \draw[dashed] (1.5,0)--(1.5,-3);
    \draw[dashed] (4.5,0)--(4.5,-3);
    \draw[<-] (1.8,-1)--(1.8,-2);
    \draw[->] (4.8,-1)--(4.8,-2);
    \draw[fermion,double] (0,-3)--(1.5,-3);
    \draw[fermion,double] (1.5,-3)--(4.5,-3);
    \draw[fermion,double] (4.5,-3)--(6,-3);
    \node at (0.75,0.6) {$p_1$};
    \node at (0.75,-3.6) {$p_2$};
    \node at (2.1,-1.5) {$l$};
    \node at (5.7,-1.5) {$l+q$};
    \end{scope}
    \begin{scope}[scale=0.55, xshift=7cm]
    \draw[fermion,double] (0,0)--(1.5,0);
    \draw[fermion,double] (1.5,0)--(4.5,0);
    \draw[fermion,double] (4.5,0)--(6,0);
    \draw[vector] (1.5,0)--(1.5,-3);
    \draw[vector] (4.5,0)--(4.5,-3);
    \draw[fermion,double] (0,-3)--(1.5,-3);
    \draw[fermion,double] (1.5,-3)--(4.5,-3);
    \draw[fermion,double] (4.5,-3)--(6,-3);
    \end{scope}
    \begin{scope}[scale=0.55, xshift=14cm]
    \draw[fermion,double] (0,0)--(1.5,0);
    \draw[fermion,double] (1.5,0)--(4.5,0);
    \draw[fermion,double] (4.5,0)--(6,0);
    \draw[dashed] (1.5,0)--(1.5,-3);
    \draw[vector] (4.5,0)--(4.5,-3);
    \draw[fermion,double] (0,-3)--(1.5,-3);
    \draw[fermion,double] (1.5,-3)--(4.5,-3);
    \draw[fermion,double] (4.5,-3)--(6,-3);
    \end{scope}
    \begin{scope}[scale=0.55, xshift=21cm]
    \draw[fermion,double] (0,0)--(1.5,0);
    \draw[fermion,double] (1.5,0)--(4.5,0);
    \draw[fermion,double] (4.5,0)--(6,0);
    \draw[vector] (1.5,0)--(1.5,-3);
    \draw[dashed] (4.5,0)--(4.5,-3);
    \draw[fermion,double] (0,-3)--(1.5,-3);
    \draw[fermion,double] (1.5,-3)--(4.5,-3);
    \draw[fermion,double] (4.5,-3)--(6,-3);
    \end{scope}
\end{tikzpicture}
    \begin{tikzpicture}
    \begin{scope}[scale=0.55]
    \draw[fermion,double] (0,0)--(1.5,0);
    \draw[fermion,double] (1.5,0)--(4.5,0);
    \draw[fermion,double] (4.5,0)--(6,0);
    \draw[dashed] (1.5,0)--(4.5,-3);
    \draw[dashed] (4.5,0)--(1.5,-3);
    \draw[<-] (1.5,-0.5)--(2.5,-1.5);
    \draw[<-] (3.5,-1.5)--(4.5,-0.5);
    \draw[fermion,double] (0,-3)--(1.5,-3);
    \draw[fermion,double] (1.5,-3)--(4.5,-3);
    \draw[fermion,double] (4.5,-3)--(6,-3);
    \node at (0.75,0.6) {$p_1$};
    \node at (0.75,-3.6) {$p_2$};
    \node at (1.6,-1.2) {$l$};
    \node at (5,-1.2) {$l+q$};
    \end{scope}
    \begin{scope}[scale=0.55, xshift=7cm]
    \draw[fermion,double] (0,0)--(1.5,0);
    \draw[fermion,double] (1.5,0)--(4.5,0);
    \draw[fermion,double] (4.5,0)--(6,0);
    \draw[vector] (1.5,0)--(4.5,-3);
    \draw[dashed] (4.5,0)--(1.5,-3);
    \draw[fermion,double] (0,-3)--(1.5,-3);
    \draw[fermion,double] (1.5,-3)--(4.5,-3);
    \draw[fermion,double] (4.5,-3)--(6,-3);
    \end{scope}
    \begin{scope}[scale=0.55, xshift=14cm]
    \draw[fermion,double] (0,0)--(1.5,0);
    \draw[fermion,double] (1.5,0)--(4.5,0);
    \draw[fermion,double] (4.5,0)--(6,0);
    \draw[dashed] (1.5,0)--(4.5,-3);
    \draw[vector] (4.5,0)--(1.5,-3);
    \draw[fermion,double] (0,-3)--(1.5,-3);
    \draw[fermion,double] (1.5,-3)--(4.5,-3);
    \draw[fermion,double] (4.5,-3)--(6,-3);
    \end{scope}
    \begin{scope}[scale=0.55, xshift=21cm]
    \draw[fermion,double] (0,0)--(1.5,0);
    \draw[fermion,double] (1.5,0)--(4.5,0);
    \draw[fermion,double] (4.5,0)--(6,0);
    \draw[vector] (1.5,0)--(4.5,-3);
    \draw[vector] (4.5,0)--(1.5,-3);
    \draw[fermion,double] (0,-3)--(1.5,-3);
    \draw[fermion,double] (1.5,-3)--(4.5,-3);
    \draw[fermion,double] (4.5,-3)--(6,-3);
    \end{scope}
\end{tikzpicture}
    \caption{Box and crossed-box diagrams contributing to the $\mathcal{O}\left(N^2\right)$ vortex-vortex potential.}
    \label{fig:box}
\end{figure}
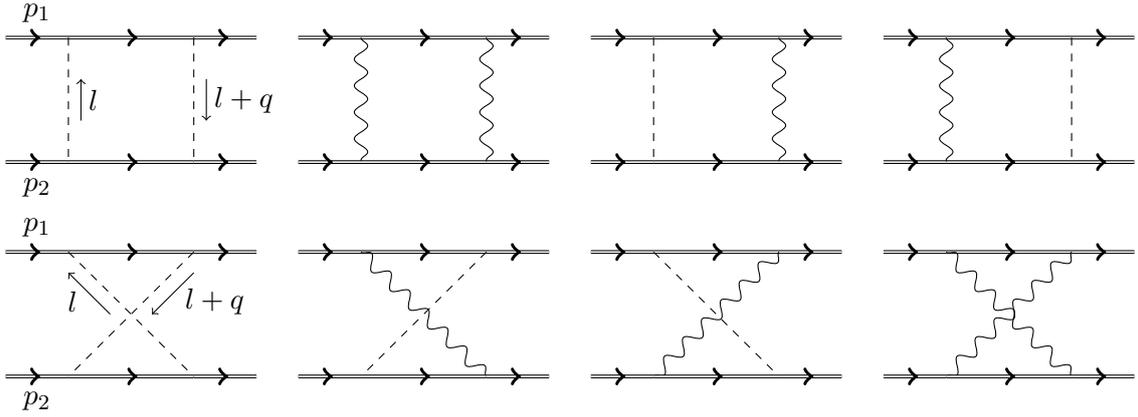

Next we consider the contribution of the box and crossed-box diagrams depicted in Figure \ref{fig:box}. For the triangle diagrams calculated above we found that the leading-order term in the soft expansion (\ref{soft}) of the one-loop amplitude scales  is $\mathcal{O}\left(q^{-2}\right)$; for the box diagrams however, the leading order term is $\mathcal{O}\left(q^{-3}\right)$, so we must expand the box diagrams to next-to-leading-order in the soft expansion. Such \textit{super-classical} terms must necessarily cancel with a corresponding iteration term to ensure a vortex-vortex potential with a smooth classical or $\frac{m}{M} \rightarrow 0$ limit. 

After reducing tensor numerators and discarding both terms without potential region contributions and short-range contributions we find, for the box diagrams
\begin{align}
    &\mathcal{M}_{\nb{}}^{(2)} \nonumber\\
    &= 256i\pi^2 M^2 N^2 \left(M^2-(p_1 \cdot p_2)\right)^2 \int \frac{\text{d}^d l}{(2\pi)^d} \frac{1}{[l^2-m^2][(l+q)^2-m^2][p_1\cdot l + i0][p_2\cdot l -i0]} \nonumber\\
    &\hspace{3mm} -128i\pi^2 m^2 M^2 N^2 \left(M^2-(p_1 \cdot p_2)\right)^2 \int \frac{\text{d}^d l}{(2\pi)^d} \frac{1}{[l^2-m^2][(l+q)^2-m^2][p_1\cdot l + i0]^2[p_2\cdot l -i0]} \nonumber\\
    &\hspace{3mm} +128i\pi^2 m^2 M^2 N^2 \left(M^2-(p_1 \cdot p_2)\right)^2 \int \frac{\text{d}^d l}{(2\pi)^d} \frac{1}{[l^2-m^2][(l+q)^2-m^2][p_1\cdot l + i0][p_2\cdot l -i0]^2} \nonumber\\
    &\hspace{3mm} -256i\pi^2 M^2 N^2 \left(M^2-(p_1 \cdot p_2)\right) \int \frac{\text{d}^d l}{(2\pi)^d} \frac{1}{[l^2-m^2][(l+q)^2-m^2][p_1\cdot l + i0]} \nonumber\\
    &\hspace{3mm} +256i\pi^2 M^2 N^2 \left(M^2-(p_1 \cdot p_2)\right) \int \frac{\text{d}^d l}{(2\pi)^d} \frac{1}{[l^2-m^2][(l+q)^2-m^2][p_2\cdot l - i0]},
\end{align}
and for the crossed-box diagrams
\begin{align}\mathcal{M}_{\cb{}}^{(2)} &=  256i\pi^2 M^2 N^2\left(M^2-(p_1\cdot p_2)\right)\int \frac{\text{d}^d l}{(2\pi)^d}\frac{1}{[l^2-m^2][(l+q)^2-m^2][p_1\cdot l +i0]}\nonumber\\
&\hspace{5mm}-256i\pi^2 M^2 N^2\left(M^2-(p_1\cdot p_2)\right)\int \frac{\text{d}^d l}{(2\pi)^d}\frac{1}{[l^2-m^2][(l+q)^2-m^2][p_2\cdot l -i0]} .
\end{align}
Interestingly, all of the super-classical contributions to the soft expansion of the crossed-box diagrams give contributions that vanish in the potential region.\footnote{Specifically they generate integrals with two linearized vortex propagators with the same sign $i0$ prescription. When expanded in the potential region, the $\omega$-integrals of such terms give zero since they are convergent at $\omega \rightarrow \infty$ and we can deform the contour in such a way that we never cross a singularity. Various subtleties in the evaluation of these $\omega$-integrals are discussed in \cite{Bern:2019crd}. } The non-vanishing crossed-box contributions exactly cancel the final two terms in the box contribution, the sum of the two therefore nicely simplifies 
\begin{align}
\label{MREFTbox}
&\mathcal{M}_{\nb{}}^{(2)}  + \mathcal{M}_{\cb{}}^{(2)}\nonumber\\
&=256i\pi^2 M^2 N^2 \left(M^2-(p_1 \cdot p_2)\right)^2 \int \frac{\text{d}^d l}{(2\pi)^d} \frac{1}{[l^2-m^2][(l+q)^2-m^2][p_1\cdot l + i0][p_2\cdot l -i0]} \nonumber\\
 &\hspace{5mm}-128i\pi^2 m^2 M^2 N^2 \left(M^2-(p_1 \cdot p_2)\right)^2 \nonumber\\
 &\hspace{5mm}\times\int \frac{\text{d}^d l}{(2\pi)^d} \frac{1}{[l^2-m^2][(l+q)^2-m^2][p_1\cdot l + i0][p_2\cdot l -i0]}\left[\frac{1}{p_1\cdot l+i0}-\frac{1}{p_2\cdot l-i0}\right].
\end{align}
We note that these contributions manifestly vanish in the static limit. In Appendix \ref{sec:soft} the expansion and resummation of these soft integrals is given. The resulting contribution to the momentum-space potential is 
\begin{align}
\label{Vbox}
&V_{\nb{}}^{(2)} + V_{\cb{}}^{(2)} \nonumber\\
 &=-\frac{128\pi^2 M^2 N^2 \mathbf{p}^4}{(\mathbf{p}^2+M^2)^{3/2}} \int \frac{\text{d}^2\mathbf{l}}{(2\pi)^2}\frac{1}{[\mathbf{l}^2+m^2][(\mathbf{l}+\mathbf{q})^2+m^2][\mathbf{p}\cdot \mathbf{l}-i0]} \nonumber\\
    &\hspace{5mm}-\frac{64\pi^2  M^2 N^2 \mathbf{p}^4}{(\mathbf{p}^2+M^2)^{3/2}} \int \frac{\text{d}^2\mathbf{l}}{(2\pi)^2}  \frac{m^2}{[\mathbf{l}^2+m^2][(\mathbf{l}+\mathbf{q})^2+m^2][\mathbf{p}\cdot \mathbf{l}-i0]^2} \nonumber\\
    &\hspace{5mm}+\frac{256  \pi^2  M N^2 \mathbf{p}^2}{\mathbf{p}^2+M^2}\left(1-\frac{M}{(\mathbf{p}^2+M^2)^{1/2}}\right) \int \frac{\text{d}^2\mathbf{l}}{(2\pi)^2} \frac{m^2}{[\mathbf{l}^2+m^2]^2[(\mathbf{l}+\mathbf{q})^2+m^2]}.
\end{align}

\subsubsection*{Iteration Contributions}

The final contribution to one-loop potential arises from the subtraction of the iteration of the tree-level potential, explicitly 
\begin{equation}
    V_{\text{iteration}}^{(2)} \equiv - \int \frac{\text{d}^{2}\mathbf{k}}{(2\pi)^{2}}\frac{V^{(1)}\left(\mathbf{p},\mathbf{k}\right)V^{(1)}\left(\mathbf{k},\mathbf{p}'\right)}{2E_{\mathbf{p}}-2E_{\mathbf{k}}+i0}.
\end{equation}
Similar to the REFT amplitude contributions, this must also be expanded in the classical potential region (\ref{probepotregion}) to sub-leading order. Using the previous result (\ref{Vtree}) and aligning the loop momenta, $\mathbf{k}= \mathbf{l}+\mathbf{p}$, we calculate 
\begin{align}
\label{Viteration}
    V^{(2)}_{\text{Iteration}} &=
     \frac{128\pi^2 M^2 N^2 \mathbf{p}^4}{\left(\mathbf{p}^2+M^2\right)^{3/2}}\int \frac{\text{d}^2\mathbf{l}}{(2\pi)^2} \frac{1}{[\mathbf{l}^2+m^2][(\mathbf{l}+\mathbf{q})^2+m^2][\mathbf{p}\cdot \mathbf{l}-i0]} \nonumber\\
     &\hspace{5mm}+\frac{64\pi^2 M^2 N^2 \mathbf{p}^4}{\left(\mathbf{p}^2+M^2\right)^{3/2}}\int \frac{\text{d}^2\mathbf{l}}{(2\pi)^2} \frac{m^2}{[\mathbf{l}^2+m^2][(\mathbf{l}+\mathbf{q})^2+m^2][\mathbf{p}\cdot \mathbf{l}-i0]^2} \nonumber\\
    &\hspace{5mm} +\frac{64\pi^2 M^2 N^2 \mathbf{p}^2\left(\mathbf{p}^2+4M^2\right)}{\left(\mathbf{p}^2+M^2\right)^{5/2}}\int \frac{\text{d}^2\mathbf{l}}{(2\pi)^2} \frac{1}{[\mathbf{l}^2+m^2][(\mathbf{l}+\mathbf{q})^2+m^2]}.
\end{align}
Importantly, not only do we find that the super-classical terms cancel in the sum of (\ref{Vbox}) and (\ref{Viteration}), so too do the classical terms proportional to the 2d integrals containing linearized propagators. This is a useful consistency check on the calculation: these terms, when integrated, generate branch cuts corresponding to the ``$s$-channel" unitarity cut separating the one-loop vortex-vortex amplitude into a pair of tree amplitudes. Since we expect that the potential is an analytic function of the momentum $\mathbf{p}$, such terms must cancel when we sum all of the contributions.

\subsubsection*{Final Result}

Combining (\ref{Vtri}), (\ref{Vbox}) and (\ref{Viteration}) together with the tree-level result (\ref{Vtree}) we find the complete $\mathcal{O}\left(N^2\right)$ vortex-vortex potential is given, in momentum space by
\begin{align}
&V^{(1)} + V^{(2)}_{\tri{}}+V^{(2)}_{\triinv{}}+V^{(2)}_{\tric{}}+V^{(2)}_{\tricinv{}} +V_{\nb{}}^{(2)} + V_{\cb{}}^{(2)} + V^{(2)}_{\text{Iteration}} \nonumber\\
&= \frac{16\pi N M \mathbf{p}^2}{\mathbf{p}^2+M^2} \frac{1}{\mathbf{q}^2+m^2}+\frac{64\pi^2 M N^2\mathbf{p}^2}{\mathbf{p}^2+M^2}\left(\frac{1}{\mathbf{q}^2+m^2}\int \frac{\text{d}^2 \mathbf{l}}{(2\pi)^2} \frac{m^2}{[\mathbf{l}^2+m^2][(\mathbf{l}+\mathbf{q})^2+m^2]}\right) \nonumber\\
   &\hspace{5mm}-\frac{64\pi^2 M N^2\mathbf{p}^2}{\mathbf{p}^2+M^2}\left(1-\frac{M(\mathbf{p}^2+4M^2)}{(\mathbf{p}^2+M^2)^{3/2}}\right)\int \frac{\text{d}^2 \mathbf{l}}{(2\pi)^2} \frac{1}{[\mathbf{l}^2+m^2][(\mathbf{l}+\mathbf{q})^2+m^2]} \nonumber\\
    &\hspace{5mm}+\frac{256  \pi^2  M N^2 \mathbf{p}^2}{\mathbf{p}^2+M^2}\left(1-\frac{M}{(\mathbf{p}^2+M^2)^{1/2}}\right) \int \frac{\text{d}^2\mathbf{l}}{(2\pi)^2} \frac{m^2}{[\mathbf{l}^2+m^2]^2[(\mathbf{l}+\mathbf{q})^2+m^2]}+\mathcal{O}\left(N^3\right).
\end{align}
To calculate the position space potential we need to Fourier transform the remaining 2d integrals.\footnote{It is certainly possible to evaluate these integrals exactly, for example the bubble integral evaluates to
\begin{equation*}
    \int \frac{\text{d}^2\mathbf{l}}{(2\pi)^2} \frac{1}{[\mathbf{l}^2+m^2][(\mathbf{l}+\mathbf{q})^2+m^2]} = \frac{\sinh ^{-1}\left(\frac{|\mathbf{q}|}{2m}\right)}{\pi   |\mathbf{q}| \sqrt{\mathbf{q}^2+4m^2}},
\end{equation*}
however, such explicit expressions make the subsequent Fourier transforms more obscure, so we will not make use of them.} This calculation is described in detail in Appendix \ref{sec:Fourier}, the final result is\\
\\
    \fbox{\begin{minipage}{\textwidth}
    \vspace{-4mm}
\begin{align}
\label{finalpotential1loop}
  V\left(\mathbf{p},\mathbf{x}\right) &=\frac{8 M \mathbf{p}^2}{\mathbf{p}^2+M^2}\left(N+\frac{2\pi}{3\sqrt{3}}N^2 \right)K_0(mr) +\frac{16M^2 N^2\mathbf{p}^2(\mathbf{p}^2+4M^2)}{(\mathbf{p}^2+M^2)^{5/2}}K_0(mr)^2 \hspace{10mm}\nonumber\\
    &\hspace{5mm}+\frac{32  M N^2 \mathbf{p}^2}{\mathbf{p}^2+M^2}\left(1-\frac{M}{(\mathbf{p}^2+M^2)^{1/2}}\right) mrK_0(mr)K_1(mr) \nonumber\\
    &\hspace{5mm}-\frac{32 M N^2\mathbf{p}^2}{\mathbf{p}^2+M^2}\left(K_0(mr)\int_{mr}^\infty \text{d}\xi \;\xi \; K_0(\xi) K_1(\xi) I_1\left(\xi\right)\right.\nonumber\\
    &\left.\hspace{35mm}+I_0\left(mr\right)  \int_{mr}^\infty \text{d}\xi\; \xi \; K_0(\xi) K_1(\xi)^2\right)+\mathcal{O}\left(N^3\right).
\end{align}
\end{minipage}}

\subsection{Moduli Space Metric}
\label{sec:moduli}

Following the well-known conjecture of Manton \cite{Manton:1981mp}, it is known that the slow-motion dynamics of critical vortices corresponds to geodesic motion on the moduli space of solutions of the BPS equations \cite{Ruback:1988ba,Samols:1991ne}. The 2-vortex moduli space metric is not known in closed form, but has been calculated numerically \cite{Shellard:1988zx,Myers:1991yh} as well as analytically at leading asymptotic order \cite{Manton:2002wb}. In this section we will derive an approximate perturbative expression for the moduli space metric from the general vortex-vortex potential (\ref{finalpotential1loop}).

To extract the metric on the moduli space we truncate the effective Hamiltonian (\ref{HNREFT}) at $\mathcal{O}(\mathbf{p}^2)$ and lift the result to a general reference frame. In general this procedure is ambiguous and naively we need to go back and recalculate the potential in a general frame. This is actually unnecessary, it is sufficient to note that the leading momentum dependence arises from the Lorentz invariant prefactors in the REFT amplitudes (\ref{MREFTtri}, \ref{MREFTbox}), which then reduce to Galilean boost invariant expressions in the non-relativistic limit, explicitly
\begin{align*}
M^2-p_1\cdot p_2 &= -\frac{1}{2}|\mathbf{p}_1-\mathbf{p}_2|^2 +\mathcal{O}\left(\mathbf{p}^4\right).
\end{align*}
From the Hamiltonian we calculate the Lagrangian by an inverse Legendre transform that we evaluate perturbatively in $N$, the result is 
\begin{equation}
    L(\dot{\mathbf{x}}_1,\mathbf{x}_1;\dot{\mathbf{x}}_2,\mathbf{x}_2) = \frac{1}{2}M \dot{\mathbf{x}}_1^2+\frac{1}{2}M \dot{\mathbf{x}}_2^2 - \tilde{U}\left(r_{12}\right)|\dot{\mathbf{x}}_1-\dot{\mathbf{x}}_2|^2,
\end{equation}
where
\begin{align}
    \tilde{U}\left(r_{12}\right)
    &= 2M\left(N+\frac{2\pi}{3\sqrt{3}}N^2\right)K_0(mr_{12})\nonumber\\
    &\hspace{5mm}-8MN^2\left(K_0(mr_{12})\int_{mr_{12}}^\infty \text{d}\xi \;\xi \; K_0(\xi) K_1(\xi) I_1\left(\xi\right)\right. \nonumber\\
    &\hspace{25mm}\left.+I_0\left(mr_{12}\right)  \int_{mr_{12}}^\infty \text{d}\xi\; \xi \; K_0(\xi) K_1(\xi)^2\right)+ \mathcal{O}(N^3).
\end{align}
According to Manton we should interpret this effective Lagrangian as defining a $0+1$d sigma model and read-off the metric on the moduli space of 2-vortices
\begin{equation}
    ds^2 = \left(\frac{1}{2}M-\tilde{U}\left(r_{12}\right)\right)d\mathbf{x}_1^2+\left(\frac{1}{2}M-\tilde{U}\left(r_{12}\right)\right)d\mathbf{x}_2^2 -2\tilde{U}\left(r_{12}\right)d\mathbf{x}_1 \cdot d\mathbf{x}_2.
\end{equation}
Isolating the leading asymptotic contribution, $\mathcal{O}\left(e^{-mr_{12}}\right)$, and using the perturbative expansion of $Z_N$ (\ref{ZNseries}) we find complete agreement with the known result \cite{Manton:2002wb}. Additionally, it is known that the moduli space metric is K\"{a}hler \cite{Samols:1991ne}. To verify this we rewrite the metric in complex coordinates 
\begin{align}
ds^2 \equiv 2g_{\alpha\overline{\beta}}dz^\alpha d\overline{z}^{\overline{\beta}} =  4M\left(dz^1 d\overline{z}^{\overline{2}}+dz^2 d\overline{z}^{\overline{1}}\right)- \tilde{U}\left(z^1,\overline{z}^{\overline{1}}\right) dz^1 d\overline{z}^{\overline{1}},
\end{align}
where
\begin{align}
z^1 &= (x_1-x_2) +i(y_1 - y_2),\hspace{5mm} \overline{z}^{\overline{1}} = (x_1-x_2) -i(y_1 - y_2) \nonumber\\
z^2 &= (x_1+x_2) +i(y_1 + y_2),\hspace{5mm} \overline{z}^{\overline{2}} = (x_1+x_2) -i(y_1 + y_2),
\end{align}
and $\mathbf{x}_i \equiv \left(x_i,y_i\right)$. The K\"{a}hler property of the metric can be straightforwardly verified from the closure of the fundamental 2-form $\Omega \equiv -2ig_{\alpha\overline{\beta}}dz^\alpha \wedge d\overline{z}^{\overline{\beta}}$ \cite{Freedman:2012zz}.\footnote{In this case closure is almost trivial and does not depend on the detailed form of $\tilde{U}(z,\overline{z})$. Rather, once the Galilean boost invariance of the $\mathcal{O}\left(\mathbf{p}^2\right)$ effective Hamiltonian is established, the K\"{a}hler property of the resulting metric is a foregone conclusion. }

\section{Discussion}
\label{sec:discussion}

The main result of this paper is (\ref{finalpotential1loop}), the $\mathcal{O}\left(N^2\right)$ contribution to the effective vortex-vortex potential. Both this result and the general point-particle EFT framework have passed non-trivial consistency checks against known results in limiting cases. In particular, the one-loop vortex-probe amplitude in the REFT (\ref{oneloopvortexprobe}) exactly matches the predicted $\mathcal{O}\left(N^{3/2}\right)$ solution of the BPS equations (\ref{pertvortexsol}) and, up to $\mathcal{O}\left(N^2 p^2\right)$ the potential (\ref{finalpotential1loop}) agrees with the prediction of the moduli space approximation at leading asymptotic order \cite{Manton:2002wb}. This gives us a great deal of confidence in the validity of this approach.

Nonetheless there are still many ways these calculations can be extended or improved. In Section \ref{sec:loop} we found that the REFT action obtained by matching probe amplitudes, with only cubic vortex-mediator interactions, was necessarily incomplete as it lead to non-vanishing static forces at $\mathcal{O}\left(N^2\right)$. We interpreted this as the requirement that the REFT must contain a ``seagull" vertex and gave a simple example (\ref{Svortex4}) that resolved the dilemma. We do not claim that this term is unique, it is also possible to engineer the cancellation of static forces by adding appropriate combinations of operators $\sigma^2|\Phi|^2$, $A_\mu A^\mu |\Phi|^2$ and $\sigma \tilde{F}_{\mu} \Phi^* \partial^\mu \Phi$. Even after imposing all relevant UV symmetries (parity, charge conjugation, $\mathcal{N}=2$ supersymmetry), there remains a one-parameter family of valid seagull terms. Stated another way, there exists an independently supersymmetric contact term that preserves all of the required symmetries and the force-cancellation condition, the coefficient of which encodes genuine UV physics and must be determined by an additional matching calculation. Physically, this Wilson coefficient encodes the linear response of the vortex to an external perturbation, equivalent to extending the matching calculation in Section \ref{sec:probetree} to $\mathcal{O}\left(\mathfrak{g}_{s,m}^2\right)$ including back-reaction effects such as the recoil of the vortex. For the analogous problem in black hole physics, various methods have been developed for calculating observables in full theory used to perform this matching, including the static response to external ``tidal" fields \cite{Hui:2020xxx} and low-energy classical Compton amplitudes \cite{Bautista:2021wfy,Bautista:2022wjf,Ivanov:2022qqt} from black hole perturbation theory, and two-body scattering observables obtained using the numerical self-force expansion \cite{Barack:2023oqp}. It would be very interesting if any of these methods could be adapted to this context. 

To obtain quantitatively accurate results for physical vortices with $N\in \mathds{Z}$, we expect to need to resum the small winding expansion as undertaken in \cite{Ohashi:2015yta} using a global Pad\'{e} approximant. This will certainly require extending the calculation of the vortex-vortex potential to higher loop order. In such a calculation the velocity resummation of soft integrals will almost certainly require the more sophisticated method of \textit{velocity differential equations} \cite{Parra-Martinez:2020dzs}. At higher-orders we also expect to encounter radiation-reaction effects. It would be particularly interesting to adapt the KMOC framework \cite{Kosower:2018adc,Herrmann:2021tct} in this context to calculate the energy loss from radiation and compare with known numerical results \cite{Myers:1991yh}. 

Finally, the calculations presented in this paper were restricted to the theoretically simpler but physically unrealistic critical case. For applications to physical systems this method should be extended to the description of vortices in the generic, non-critical AHM. This should be possible, at least in the weakly type-I or type-II regimes where departure from criticality can be treated as a perturbation. We expect the same separation of scales should occur in the small winding limit and so the same point-particle EFT methods should apply. We leave this and other interesting generalizations to future work. 

\vspace{3mm}
\noindent \textbf{Acknowledgements}

\vspace{3mm}
I would like to thank Zvi Bern, Thomas Dumitrescu, Amey Gaikwad, Shruti Paranjape and Michael Ruf for useful discussions. CRTJ is supported by the Department of Energy under Award Number DE-SC0009937 and acknowledges the continued support of the Mani L. Bhaumik Institute for Theoretical Physics. 

\appendix

\section{Conventions}
\label{sec:conventions}

In this paper we work in the mostly minus metric convention
\begin{equation}
    \eta_{\mu\nu} = \eta^{\mu\nu} = 
    \begin{pmatrix}
        1 & 0 & 0 \\
        0 & -1 & 0 \\
        0 & 0 & -1
    \end{pmatrix}.
\end{equation}
The Levi-Civita symbol in $2+1d$ is defined by the sign convention $\epsilon^{012} = \epsilon_{012} = +1$, while the $2d$ (spatial) Levi-Civita is defined in the convention $\epsilon^{12} = +1$. We choose to work with the following basis of real and symmetric Pauli matrices 
\begin{align}
    &\sigma_{ab}^0 = \begin{pmatrix}1 & 0 \\ 0& 1\end{pmatrix}, \hspace{10mm} \sigma_{ab}^1 = \begin{pmatrix}0 & 1 \\ 1& 0\end{pmatrix}, \hspace{10mm} \sigma_{ab}^2 = \begin{pmatrix}1 & 0 \\ 0& -1\end{pmatrix}.
\end{align}
In $d=2+1$ there is only one type of spinor index which can be raised/lowered with the spinor Levi-Civita symbol
\begin{equation}
    \epsilon^{ac}\epsilon^{bd}\sigma^\mu_{cd} \equiv \overline{\sigma}^{\mu ab},
\end{equation}
defined in the sign convention $\epsilon^{12}=-\epsilon_{12}=+1$. The Pauli matrices satisfy the usual Clifford algebra relation
\begin{equation}
    \sigma^\mu_{ac} \overline{\sigma}^{\nu cb} + \sigma^\nu_{ac} \overline{\sigma}^{\mu cb} = 2\eta^{\mu\nu} \delta^b_a.
\end{equation}
Our conventions for the $\mathcal{N}=2$ super-Poincare algebra in $d=2+1$ are 
\begin{align}
    \{Q_a,Q^\dagger_b\} &= 2\sigma^\mu_{ab}P_\mu - 2iZ \epsilon_{ab}\nonumber\\
    [Q_a,M^{\mu\nu}] &= \frac{i}{4}{\left(\sigma^{\mu}\overline{\sigma}^\nu - \sigma^{\nu}\overline{\sigma}^\mu\right)_a}^b Q_b \nonumber\\
    [Q^\dagger_a,M^{\mu\nu}] &= \frac{i}{4}{\left(\sigma^{\mu}\overline{\sigma}^\nu - \sigma^{\nu}\overline{\sigma}^\mu\right)_a}^b Q^\dagger_b \nonumber\\
    [M^{\mu\nu},P^\rho] &= -i\left(\eta^{\mu\rho}P^\nu - \eta^{\nu \rho} P^\mu\right) \nonumber\\
    [M^{\mu\nu},M^{\rho\sigma}] &= -i\left(\eta^{\mu\rho}M^{\nu\sigma}- \eta^{\nu\rho}M^{\mu\sigma} -\eta^{\mu\sigma}M^{\nu\rho}+\eta^{\nu\sigma}M^{\mu\rho}\right),
\end{align}
with all other (anti-)commutators vanishing.

\subsubsection*{Massive Spinor Variables}

To construct an on-shell superspace in Section \ref{sec:susy} we make use of massive spinor variables in $d=2+1$ defined by the conditions
\begin{equation}
    p_{ab} = p_\mu \sigma^\mu_{ab} = \lambda_a \tilde{\lambda}_b + \lambda_b \tilde{\lambda}_a, \hspace{10mm} \lambda_a \tilde{\lambda}^a = im,
\end{equation}
where $p^2=m^2$ and indices on the spinors are raised/lowered with the 2d Levi-Civita symbol
\begin{equation}
    \lambda^a = \epsilon^{ab}\lambda_b, \hspace{10mm} \tilde{\lambda}^a = \epsilon^{ab} \tilde{\lambda}_b.
\end{equation}
There is a redundancy in this description 
\begin{equation}
    \lambda_a \rightarrow e^{i\theta} \lambda_a,\hspace{10mm} \tilde{\lambda}_a \rightarrow e^{-i\theta} \tilde{\lambda}_a,
\end{equation}
corresponding to the action of the little group of massive particles $SO(2)\cong U(1)$. The Lorentz invariant spin quantum number labelling little group representations corresponds to the eigenvalue of the Pauli-Lubanski scalar 
\begin{equation}
    W = -\frac{1}{2}\epsilon_{\mu\nu\rho}P^\mu M^{\nu\rho}.
\end{equation}
We will work in a convention in which $\lambda_a$ carries $+1/2$ units of spin and $\tilde{\lambda}_a$ carries $-1/2$ units. An explicit solution to these conditions for real-valued momenta with $p^0>0$ are given by
\begin{equation}
    \lambda_a = \frac{1}{\sqrt{2\left(p^0+p^2\right)}}\begin{pmatrix}
        p^1-im \\
        -p^0 -p^2
    \end{pmatrix}, \hspace{5mm}\tilde{\lambda}_a = \frac{1}{\sqrt{2\left(p^0+p^2\right)}}\begin{pmatrix}
        p^1+im \\
        -p^0 -p^2
    \end{pmatrix}.
\end{equation}
From these expressions we see that these spinors are interchanged under crossing up to a possible phase factor, a consistent choice is given by
\begin{equation}
    \lambda_{-p}^a = i\tilde{\lambda}_p^a, \hspace{5mm} \tilde{\lambda}_{-p}^a = i\lambda_p^a.
\end{equation}
It is convenient to define Lorentz invariant spinor brackets, we will use 
\begin{equation}
    \langle pq \rangle = \tilde{\lambda}_{pa} \tilde{\lambda}_{q}^a, \hspace{10mm} [pq] = \lambda_{pa} \lambda_{q}^a, \hspace{10mm} \langle pq ]= \tilde{\lambda}_{pa} \lambda_{q}^a, \hspace{10mm} [ pq \rangle= \lambda_{pa} \tilde{\lambda}_{q}^a.
\end{equation}
Many Schouten-type identities among the spinor brackets can be derived from the fundamental identity
\begin{equation}
    \epsilon^{ab}\epsilon^{cd} + \epsilon^{ac}\epsilon^{db} + \epsilon^{ad}\epsilon^{bc} = 0.
\end{equation}
Finally we can derive expressions for the polarization vectors of an (outgoing) massive spin-1 particle
\begin{equation}
    \varepsilon_+^\mu(p) = \frac{1}{\sqrt{2}m}\lambda_a \overline{\sigma}^{\mu ab}\lambda_b, \hspace{10mm} \varepsilon_-^\mu(p) = \frac{1}{\sqrt{2}m}\tilde{\lambda}_a \overline{\sigma}^{\mu ab}\tilde{\lambda}_b.
\end{equation}
Using the above results we derive the crossing 
\begin{equation}
    \varepsilon_{\pm}^\mu(-p) = -\varepsilon_{\mp}^\mu(p),
\end{equation}
and completeness relations
\begin{equation}
    \sum_{\pm} \varepsilon_{\pm}^\mu(-p) \varepsilon_{\pm}^\nu(p) = \eta^{\mu\nu} - \frac{p^\mu p^\nu}{m^2}.
\end{equation}

\section{Feynman Rules}
\label{sec:Feynman}

\subsubsection*{Critical Abelian Higgs Model}

The Feynman rules for the self-interactions of the mediator particles in the critical AHM, in unitary gauge and in the notation corresponding to (\ref{AHMmanifestcounting}) are given by:
\begin{center}
    \begin{tikzpicture}
        \draw[-,dashed] (0,0)--(2,0);
        \draw[->] (0.5,0.3)--(1.5,0.3);
        \node at (1,0.6) {$k$};
        \node at (5,0.2) {
        \begin{minipage}{3cm}
            \begin{equation*}
                = \hspace{12mm}\frac{i}{k^2-m^2+i0}
            \end{equation*}
        \end{minipage}
        };
    \end{tikzpicture}
\end{center}
\vspace{5mm}
\begin{center}
    \begin{tikzpicture}
        \draw[-,decorate,decoration=snake] (0,0)--(2,0);
        \draw[->] (0.5,0.3)--(1.5,0.3);
        \node at (1,0.6) {$k$};
        \node at (-0.2,0.2) {$\mu$};
        \node at (2.2,0.2) {$\nu$};
        \node at (5,0.2) {
        \begin{minipage}{3cm}
            \begin{equation*}
                = \hspace{12mm}\frac{-i\left(\eta^{\mu\nu}-\frac{k^\mu k^\nu}{m^2}\right)}{k^2-m^2+i0}
            \end{equation*}
        \end{minipage}
        };
        \node at (7,0) {};
    \end{tikzpicture}
\end{center}
\vspace{5mm}
\begin{center}
    \begin{tikzpicture}
        \draw[-,dashed] (0,0)--(1.5,0);
        \draw[-,decorate,decoration=snake] (1.5,0)--(2.25,1.3);
        \draw[-,decorate,decoration=snake] (1.5,0)--(2.25,-1.3);
        \node at (2.4,1.5) {$\mu$};
        \node at (2.4,-1.5) {$\nu$};
        \node at (5,0.2) {
        \begin{minipage}{3cm}
            \begin{equation*}
                = \hspace{12mm}i\sqrt{2\pi}M^{3/2} \left(\frac{m}{M}\right)^2 N^{1/2}\eta^{\mu\nu}
            \end{equation*}
        \end{minipage}
        };
    \end{tikzpicture}
\end{center}
\vspace{5mm}
\begin{center}
    \begin{tikzpicture}
        \draw[-,dashed] (0,0)--(1.5,0);
        \draw[-,dashed] (1.5,0)--(2.25,1.3);
        \draw[-,dashed] (1.5,0)--(2.25,-1.3);
        \node at (5,0.2) {
        \begin{minipage}{3cm}
            \begin{equation*}
                = \hspace{12mm}-3i\sqrt{\frac{\pi}{2}} M^{3/2}\left(\frac{m}{M}\right)^2 N^{1/2}
            \end{equation*}
        \end{minipage}
        };
    \end{tikzpicture}
\end{center}
\vspace{5mm}
\begin{center}
    \begin{tikzpicture}
        \draw[-,dashed] (-1,1)--(0,0);
        \draw[-,dashed] (-1,-1)--(0,0);
        \draw[-,dashed] (1,1)--(0,0);
        \draw[-,dashed] (1,-1)--(0,0);
        \node at (4,0.2) {
        \begin{minipage}{3cm}
            \begin{equation*}
                = \hspace{12mm} -\frac{3i}{2} M \left(\frac{m}{M}\right)^2 N
            \end{equation*}
        \end{minipage}
        };
    \end{tikzpicture}
\end{center}
\vspace{5mm}
\begin{center}
    \begin{tikzpicture}
        \draw[-,decorate,decoration=snake] (-1,1)--(0,0);
        \draw[-,decorate,decoration=snake] (-1,-1)--(0,0);
        \draw[-,dashed] (1,1)--(0,0);
        \draw[-,dashed] (1,-1)--(0,0);
        \node at (-1.2,1.2) {$\mu$};
        \node at (-1.2,-1.2) {$\nu$};
        \node at (4,0.2) {
        \begin{minipage}{3cm}
            \begin{equation*}
                = \hspace{12mm}i\pi M \left(\frac{m}{M}\right)^2 N \eta^{\mu\nu}.
            \end{equation*}
        \end{minipage}
        };
    \end{tikzpicture}
\end{center}

\subsubsection*{Relativistic EFT}

The Feynman rules for the interactions between the (point-particle) vortex and the mediator particles corresponding to the REFT effective action (\ref{Svortex2}), (\ref{Svortex3}) and the scalar-scalar interaction in (\ref{Svortex4}) are given by:

\begin{center}
    \begin{tikzpicture}
        \draw[fermion,double] (0,0)--(2,0);
        \draw[->] (0.5,0.3)--(1.5,0.3);
        \node at (1,0.6) {$k$};
        \node at (5,0.2) {
        \begin{minipage}{3cm}
            \begin{equation*}
                = \hspace{12mm}\frac{i}{k^2-M^2+i0}
            \end{equation*}
        \end{minipage}
        };
    \end{tikzpicture}
\end{center}
\vspace{5mm}
\begin{center}
    \begin{tikzpicture}
        \draw[-,dashed] (0,0)--(1.5,0);
        \draw[fermion,double] (1.5,0)--(2.25,1.3);
        \draw[fermionbar,double] (1.5,0)--(2.25,-1.3);
        \node at (5,0.2) {
        \begin{minipage}{3cm}
            \begin{equation*}
                = \hspace{12mm}ig_s M^{3/2}N^{1/2}
            \end{equation*}
        \end{minipage}
        };
        \node at (5,0) {};
    \end{tikzpicture}
\end{center}
\vspace{5mm}
\begin{center}
    \begin{tikzpicture}
        \draw[-,decorate,decoration=snake] (0,0)--(1.5,0);
        \draw[fermion,double] (1.5,0)--(2.25,1.3);
        \draw[fermionbar,double] (1.5,0)--(2.25,-1.3);
        \draw[->] (1.8,-1.2)--(1.35,-0.4);
        \draw[->] (1.2,0.3)--(0.4,0.3);
        \node at (0.75,0.6) {$q$};
        \node at (1.2,-1) {$p$};
        \node at (-0.3,0) {$\mu$};
        \node at (5.25,0.2) {
        \begin{minipage}{3cm}
            \begin{equation*}
                = \hspace{12mm}-4g_m M^{-1/2}N^{1/2}\left(\frac{m}{M}\right)^{-1} \epsilon^{\mu\nu\rho}q_\nu p_\rho
            \end{equation*}
        \end{minipage}
        };
        \node at (7.5,0) {};
    \end{tikzpicture}
\end{center}
\vspace{5mm}
\begin{center}
        \begin{tikzpicture}
        \draw[-,dashed] (-1,1)--(0,0);
        \draw[-,dashed] (-1,-1)--(0,0);
        \draw[fermionbar,double] (1,1)--(0,0);
        \draw[fermion,double] (1,-1)--(0,0);
        \node at (4,0.2) {
        \begin{minipage}{3cm}
            \begin{equation*}
                = \hspace{12mm} 2ig_{\sigma\sigma}M N
            \end{equation*}
        \end{minipage}
        };
    \end{tikzpicture}
\end{center}
The cubic vortex-mediator couplings were determined in (\ref{probematchingresult}) by matching with full theory and the quartic coupling fixed in (\ref{gss}) by requiring the cancellation of static forces at one-loop.\\
\\
In addition we also need the Feynman rules for the probe particle defined by the effective action (\ref{Sprobe}), these are given by:

\begin{center}
    \begin{tikzpicture}
        \draw[fermion] (0,0)--(2,0);
        \draw[->] (0.5,0.3)--(1.5,0.3);
        \node at (1,0.6) {$k$};
        \node at (5,0.2) {
        \begin{minipage}{3cm}
            \begin{equation*}
                = \hspace{12mm}\frac{i}{k^2-\mathfrak{m}^2+i0}
            \end{equation*}
        \end{minipage}
        };
    \end{tikzpicture}
\end{center}
\vspace{5mm}
\begin{center}
    \begin{tikzpicture}
        \draw[-,dashed] (0,0)--(1.5,0);
        \draw[fermion] (1.5,0)--(2.25,1.3);
        \draw[fermionbar] (1.5,0)--(2.25,-1.3);
        \node at (5,0.2) {
        \begin{minipage}{3cm}
            \begin{equation*}
                = \hspace{12mm}i\mathfrak{g}_s 
            \end{equation*}
        \end{minipage}
        };
        \node at (7.5,0) {};
    \end{tikzpicture}
\end{center}
\vspace{5mm}
\begin{center}
    \begin{tikzpicture}
        \draw[-,decorate,decoration=snake] (0,0)--(1.5,0);
        \draw[fermion] (1.5,0)--(2.25,1.3);
        \draw[fermionbar] (1.5,0)--(2.25,-1.3);
        \draw[->] (1.8,-1.2)--(1.35,-0.4);
        \draw[->] (1.2,0.3)--(0.4,0.3);
        \node at (0.75,0.6) {$q$};
        \node at (1.2,-1) {$p$};
        \node at (-0.3,0) {$\mu$};
        \node at (5.25,0.2) {
        \begin{minipage}{3cm}
            \begin{equation*}
                = \hspace{12mm}-4\mathfrak{g}_m \epsilon^{\mu\nu\rho}q_\nu p_\rho.
            \end{equation*}
        \end{minipage}
        };
        \node at (7.5,0) {};
    \end{tikzpicture}
\end{center}

\subsubsection*{Non-Relativistic EFT}

The Feynman rules derived from the NREFT effective action (\ref{SNREFT}) are given by:
\begin{center}
    \begin{tikzpicture}
        \draw[-,fermion,double] (-1.5,0)--(1.5,0);
        \node at (0,0.4) {$\left(\omega,\mathbf{k}\right)$};
        \node at (4,0.2) {
        \begin{minipage}{3cm}
            \begin{equation*}
                = \hspace{12mm} \frac{i}{\omega-\sqrt{\mathbf{k}^2+M^2}+i0}
            \end{equation*}
        \end{minipage}
        };
        \node at (6,0) {};
    \end{tikzpicture}
\end{center}
\vspace{5mm}
\begin{center}
    \begin{tikzpicture}
        \draw[-,fermion,double] (-1,1)--(0,0);
        \draw[-,fermion,double] (-1,-1)--(0,0);
        \draw[fermionbar,double] (1,1)--(0,0);
        \draw[fermionbar,double] (1,-1)--(0,0);
        \node at (4,0.2) {
        \begin{minipage}{3cm}
            \begin{equation*}
                = \hspace{12mm} -iV\left(\mathbf{k},\mathbf{k}'\right).
            \end{equation*}
        \end{minipage}
        };
        \node at (-0.5,0.9) {$\mathbf{k}$};
        \node at (-0.5,-0.9) {$-\mathbf{k}$};
        \node at (0.5,0.9) {$\mathbf{k}'$};
        \node at (0.4,-0.9) {$-\mathbf{k}'$};
    \end{tikzpicture}
\end{center}

\section{Velocity Resummation of Soft Integrals}
\label{sec:soft}

\subsubsection*{Triangle Integral}

The triangle soft integral is defined as
\begin{equation}
    \mathcal{I}_{\tri{}} = \int \frac{\text{d}^d l}{(2\pi)^d} \frac{1}{[l^2-m^2][(l+q)^2-m^2][p_1\cdot l + i0]}.
\end{equation}
To simplify this we first note that if we make a change of variables $l\rightarrow -l-q$ and drop a quantum term (which must vanish in the final result since the original integral was homogeneous) we find 
\begin{equation}
    -\int \frac{\text{d}^d l}{(2\pi)^d} \frac{1}{[l^2-m^2][(l+q)^2-m^2][p_1\cdot l -i0]}.
\end{equation}
This is not exactly the same as the original integral since the $i0$ in the linearized propagator has changed sign. Using this we can rewrite the original integral as 
\begin{align}
    \mathcal{I}_{\tri{}} &= \frac{1}{2}\int \frac{\text{d}^d l}{(2\pi)^d} \frac{1}{[l^2-m^2][(l+q)^2-m^2]}\left(\frac{1}{p_1\cdot l + i0}- \frac{1}{p_1\cdot l - i0}\right) \nonumber\\
    &= -i\pi\int \frac{\text{d}^d l}{(2\pi)^d} \frac{\delta\left(p_1\cdot l\right)}{[l^2-m^2][(l+q)^2-m^2]},
\end{align}
where we have used the distributional identity
\begin{equation}
    \frac{1}{x+i0}-\frac{1}{x-i0} = -2\pi i\delta(x).
\end{equation}
In this form we can calculate the $\omega$-integral explicitly 
\begin{align}
    &= -i\pi\int \frac{\text{d}^2 \mathbf{l}}{(2\pi)^2} \frac{\text{d}\omega}{2\pi} \frac{\delta\left(E\omega-\mathbf{p}\cdot \mathbf{l}\right)}{[\omega^2-\mathbf{l}^2-m^2][\omega^2 -(\mathbf{l}+\mathbf{q})^2-m^2]} \nonumber\\
    &= -\frac{i}{2E}\int \frac{\text{d}^2 \mathbf{l}}{(2\pi)^2} \frac{1}{[(\mathbf{u}\cdot \mathbf{l})^2-\mathbf{l}^2-m^2][(\mathbf{u}\cdot \mathbf{l})^2 -(\mathbf{l}+\mathbf{q})^2-m^2]},
\end{align}
where we have defined $\mathbf{u}=\frac{\mathbf{p}}{E}$. Next we expand the integrand order-by-order in $\mathbf{u}$ and simplify the resulting tensor integrals using IBP relations. It is convenient to introduce some notation, define
\begin{equation}
    I_{\alpha,\beta}^{(n)} \equiv \int \frac{\text{d}^2 \mathbf{l}}{(2\pi)^2} \frac{(\mathbf{u}\cdot\mathbf{l})^n}{[\mathbf{l}^2+m^2]^\alpha [(\mathbf{l}+\mathbf{q})^2+m^2]^\beta}.
\end{equation}
By making a change of variables $\mathbf{l}\rightarrow-\mathbf{l}-\mathbf{q}$ and dropping a quantum term we can see that these satisfy the symmetry property $I_{\alpha,\beta}^{(n)} = (-1)^n I_{\beta,\alpha}^{(n)}$. More non-trivially from the IBP relation
\begin{equation}
    \int \frac{\text{d}^2 \mathbf{l}}{(2\pi)^2} \; \mathbf{u}\cdot \frac{\partial}{\partial \mathbf{l}}\left(\frac{(\mathbf{u}\cdot\mathbf{l})^n}{[\mathbf{l}^2+m^2]^\alpha [(\mathbf{l}+\mathbf{q})^2+m^2]^\beta}\right) = 0,
\end{equation}
we derive the identity
\begin{equation}
    \label{IBP}
    n \mathbf{u}^2 I_{\alpha,\beta}^{(n-1)} = 2\alpha I_{\alpha+1,\beta}^{(n+1)}+2\beta I_{\alpha,\beta+1}^{(n+1)}.
\end{equation}
This turns out to be all we need to simplify the above integral. It is straightforward to calculate 
\begin{align}
    &-\frac{i}{2E}\int \frac{\text{d}^2 \mathbf{l}}{(2\pi)^2} \frac{1}{[(\mathbf{u}\cdot \mathbf{l})^2-\mathbf{l}^2-m^2][(\mathbf{u}\cdot \mathbf{l})^2 -(\mathbf{l}+\mathbf{q})^2-m^2]} \nonumber\\
    &= -\frac{i}{2E}\left(1+\frac{\mathbf{u}^2}{2}+\frac{3\mathbf{u}^4}{4}+\frac{5\mathbf{u}^6}{16}+\frac{35\mathbf{u}^8}{128}+...\right)\int \frac{\text{d}^2 \mathbf{l}}{(2\pi)^2} \frac{1}{[\mathbf{l}^2+m^2][(\mathbf{l}+\mathbf{q})^2+m^2]}.
\end{align}
This is enough to recognize the series and resum
\begin{equation}
    = -\frac{i}{2E\sqrt{1-\mathbf{u}^2}}\int \frac{\text{d}^2 \mathbf{l}}{(2\pi)^2} \frac{1}{[\mathbf{l}^2+m^2][(\mathbf{l}+\mathbf{q})^2+m^2]},
\end{equation}
rewriting in terms of the original variables we have the resummed triangle integral
\begin{align}
    \mathcal{I}_{\tri{}} &= \int \frac{\text{d}^d l}{(2\pi)^d} \frac{1}{[l^2-m^2][(l+q)^2-m^2][p_1\cdot l + i0]} \nonumber\\
        &= -\frac{i}{2M}\int \frac{\text{d}^2 \mathbf{l}}{(2\pi)^2} \frac{1}{[\mathbf{l}^2+m^2][(\mathbf{l}+\mathbf{q})^2+m^2]}.
\end{align}

\subsubsection*{(Super-classical) Box Integral}

The super-classical box soft integral is defined as
\begin{equation}
    \mathcal{I}_{\nb{}}^{(\text{sc})} = \int \frac{\text{d}^d l}{(2\pi)^d} \frac{1}{[l^2-m^2][(l+q)^2-m^2][p_1\cdot l + i0][p_2\cdot l -i0]}.
\end{equation}
Rewriting this in non-relativistic notation 
\begin{equation}
    = \frac{1}{E^2}\int \frac{\text{d}^2\mathbf{l}}{(2\pi)^2}\frac{\text{d}\omega}{2\pi} \frac{1}{[\omega^2-\mathbf{l}^2-m^2][\omega^2-(\mathbf{l}+\mathbf{q})^2-m^2][\omega-\mathbf{u}\cdot \mathbf{l}+i0][\omega+\mathbf{u}\cdot \mathbf{l}-i0]}.
\end{equation}
To evaluate this as a series in $\mathbf{u}$ we will expand in a region defined by the scaling 
\begin{equation}
    \omega \sim v, \hspace{5mm} \mathbf{u} \sim v.
\end{equation}
This is morally the same as the potential region except that we are keeping the velocity dependency in the $E$ factors implicit in $\mathbf{u}$ intact. Order-by-order in this expansion we encounter $\omega$-integrals of the form
\begin{equation}
    \int \frac{\text{d}\omega}{2\pi} \frac{\omega^n}{[\omega-\mathbf{u}\cdot \mathbf{l}+i0][\omega+\mathbf{u}\cdot \mathbf{l}-i0]} = \frac{i(-1)^{n-1}}{2}\left(\mathbf{u}\cdot \mathbf{l}-i0\right)^{n-1}.
\end{equation}
These have been evaluated by analytic continuation from the range $-1<n<1$ where the integral is convergent. Now we note that since the (non-linearized) propagators in the original integral are functions of $\omega^2$, we only need $\omega$-integrals of this kind for $n$ an even, non-negative integer. For $n>0$ we can ignore the $i0$ since it appears in the numerator and we are left to simplify 2d integrals of the form 
\begin{equation}
    \int \frac{\text{d}^2\mathbf{l}}{(2\pi)^2} \frac{(\mathbf{u}\cdot \mathbf{l})^{2(n_1+n_2)-1}}{[\mathbf{l}^2+m^2]^{n_1}[(\mathbf{l}+\mathbf{q})^2+m^2]^{n_2}}.
\end{equation}
All such integrals appear in pairs together with $n_1\leftrightarrow n_2$, and the sum of the two vanish since the change of variables $\mathbf{l}\rightarrow -\mathbf{l}-\mathbf{q}$ changes the overall sign. For the $n=0$ term, corresponding to expanding both mediator propagators at leading order, we can no longer ignore the $i0$ and so the same trick doesn't apply. This term is non-zero and gives the complete super-classical box soft integral
\begin{align}
    \mathcal{I}_{\nb{}}^{(\text{sc})} &= \int \frac{\text{d}^d l}{(2\pi)^d} \frac{1}{[l^2-m^2][(l+q)^2-m^2][p_1\cdot l + i0][p_2\cdot l -i0]} \nonumber\\
    &= -\frac{i}{2E}\int \frac{\text{d}^2\mathbf{l}}{(2\pi)^2}\frac{1}{[\mathbf{l}^2+m^2][(\mathbf{l}+\mathbf{q})^2+m^2][\mathbf{p}\cdot \mathbf{l}-i0]}.
\end{align}

\subsubsection*{(Classical) Box Integral}

The classical box soft integral is defined as 
\begin{equation}
    \mathcal{I}_{\nb{}}^{(\text{c})} = \int \frac{\text{d}^d l}{(2\pi)^d} \frac{1}{[l^2-m^2][(l+q)^2-m^2][p_1\cdot l + i0][p_2\cdot l -i0]}\left[\frac{1}{p_1\cdot l+i0}-\frac{1}{p_2\cdot l-i0}\right].
\end{equation}
We calculate this using a combination of the methods used above, begin by rewriting it in non-relativistic notation
\begin{align}
    &= \frac{1}{E^3}\int \frac{\text{d}^2\mathbf{l}}{(2\pi)^2} \frac{\text{d} \omega}{2\pi} \frac{1}{[\omega^2-\mathbf{l}^2-m^2][\omega^2-(\mathbf{l}+\mathbf{q})^2-m^2][\omega -\mathbf{u}\cdot \mathbf{l}+i0][\omega +\mathbf{u}\cdot \mathbf{l}-i0]}\nonumber\\
    &\hspace{30mm}\times\left[\frac{1}{\omega -\mathbf{u}\cdot \mathbf{l}+i0}-\frac{1}{\omega +\mathbf{u}\cdot \mathbf{l}-i0}\right].
\end{align}
We then expand the integrand in $\mathbf{u}$ in the potential region as we did above for the super-classical box and evaluate the $\omega$-integrals by analytic continuation. At $\mathcal{O}(\omega^0)$ we find a term with a spurious matter propagator
\begin{equation}
    \mathcal{I}_{\nb{}}^{(\text{c})}\vert_{\omega^0} = \frac{i}{2E^3} \int \frac{\text{d}^2\mathbf{l}}{(2\pi)^2}  \frac{1}{[\mathbf{l}^2+m^2][(\mathbf{l}+\mathbf{q})^2+m^2][\mathbf{u}\cdot \mathbf{l}-i0]^2}.
\end{equation}
At higher-orders in the expansion we find only powers of $(\mathbf{u}\cdot \mathbf{l})$ in the numerator, these can then be simplified using the IBP relations (\ref{IBP}) with the resummed result 
\begin{align}
    \mathcal{I}_{\nb{}}^{(\text{c})} &= \int \frac{\text{d}^d l}{(2\pi)^d} \frac{1}{[l^2-m^2][(l+q)^2-m^2][p_1\cdot l + i0][p_2\cdot l -i0]}\left[\frac{1}{p_1\cdot l+i0}-\frac{1}{p_2\cdot l-i0}\right] \nonumber\\
    &=\frac{i}{2E^3} \int \frac{\text{d}^2\mathbf{l}}{(2\pi)^2}  \frac{1}{[\mathbf{l}^2+m^2][(\mathbf{l}+\mathbf{q})^2+m^2][\mathbf{u}\cdot \mathbf{l}-i0]^2} \nonumber\\
    &\hspace{5mm}-\frac{i}{E^3}\left(\frac{1}{2}+\frac{3\mathbf{u}^2}{4}+ \frac{5\mathbf{u}^4}{8}+ \frac{35\mathbf{u}^6}{64}+ \frac{63\mathbf{u}^8}{128}+ \frac{231\mathbf{u}^{10}}{512}+...\right)\nonumber\\
    &\hspace{20mm}\times\int \frac{\text{d}^2\mathbf{l}}{(2\pi)^2} \frac{1}{[\mathbf{l}^2+m^2]^2[(\mathbf{l}+\mathbf{q})^2+m^2]} \nonumber\\
    &=\frac{i}{2E} \int \frac{\text{d}^2\mathbf{l}}{(2\pi)^2}  \frac{1}{[\mathbf{l}^2+m^2][(\mathbf{l}+\mathbf{q})^2+m^2][\mathbf{p}\cdot \mathbf{l}-i0]^2} \nonumber\\
    &\hspace{10mm}- \frac{2i}{ME (M+E)}\int \frac{\text{d}^2\mathbf{l}}{(2\pi)^2} \frac{1}{[\mathbf{l}^2+m^2]^2[(\mathbf{l}+\mathbf{q})^2+m^2]}.
\end{align}
It is possible to use an IBP relation to rewrite the integral in the second term as
\begin{equation}
    \int \frac{\text{d}^2\mathbf{l}}{(2\pi)^2} \frac{1}{[\mathbf{l}^2+m^2]^2[(\mathbf{l}+\mathbf{q})^2+m^2]} = \frac{1}{\mathbf{q}^2+4m^2}\left(\frac{1}{4\pi m^2}+ \int \frac{\text{d}^2\mathbf{l}}{(2\pi)^2} \frac{1}{[\mathbf{l}^2+m^2][(\mathbf{l}+\mathbf{q})^2+m^2]}\right).
\end{equation}
This actually makes the subsequent Fourier transform more complicated so we will leave our result for the classical box soft integral in the form above.

\section{Fourier Transforms}
\label{sec:Fourier}

To obtain a position space potential we need to Fourier transform the 2d integrals above. Since the integrals with linearized matter poles have to cancel with the iteration contributions we will not evaluate these. Two basic results we will need are 
\begin{align}
    \int \frac{\text{d}^2\mathbf{q}}{(2\pi)^2} \frac{e^{i\mathbf{q}\cdot \mathbf{x}}}{\mathbf{q}^2+m^2} &= \frac{1}{2\pi}K_0\left(mr\right) \nonumber\\
    \int \frac{\text{d}^2\mathbf{q}}{(2\pi)^2} \frac{e^{i\mathbf{q}\cdot \mathbf{x}}}{[\mathbf{q}^2+m^2]^2} &= \frac{r}{4\pi m} K_1(mr).
\end{align}
The first is enough to calculate the tree-level potential. At one-loop the functions we are Fourier transforming are Feynman integrals. Even though these integrals can be calculated explicitly in momentum space, as we will see, it is simpler to Fourier transform before loop integration. For one-loop integrals with ``\Large$\tric{}$\normalsize"\;topology the Fourier transform is very simple
\begin{align}
    &\int \frac{\text{d}^2 \mathbf{q}}{(2\pi)^2} e^{i\mathbf{q}\cdot \mathbf{x}} \int \frac{\text{d}^2\mathbf{l}}{(2\pi)^2} \frac{1}{[\mathbf{l}^2+m^2][(\mathbf{l}+\mathbf{q})^2+m^2]} \nonumber\\
    &= \int \frac{\text{d}^2 \mathbf{q}}{(2\pi)^2} \int \frac{\text{d}^2\mathbf{l}}{(2\pi)^2} \frac{e^{i(\mathbf{q}-\mathbf{l})\cdot \mathbf{x}} }{[\mathbf{l}^2+m^2][\mathbf{q}^2+m^2]} \nonumber\\
    &= \left[\int \frac{\text{d}^2\mathbf{q}}{(2\pi)^2} \frac{e^{i\mathbf{q}\cdot \mathbf{x}}}{\mathbf{q}^2+m^2}\right]\left[\int \frac{\text{d}^2\mathbf{l}}{(2\pi)^2} \frac{e^{-i\mathbf{l}\cdot \mathbf{x}}}{\mathbf{l}^2+m^2}\right].
\end{align}
In this factored form we can make use of the results above 
\begin{equation}
    \int \frac{\text{d}^2 \mathbf{q}}{(2\pi)^2} e^{i\mathbf{q}\cdot \mathbf{x}} \int \frac{\text{d}^2\mathbf{l}}{(2\pi)^2} \frac{1}{[\mathbf{l}^2+m^2][(\mathbf{l}+\mathbf{q})^2+m^2]} = \frac{1}{4\pi^2} K_0(mr)^2.
\end{equation}
Similarly 
\begin{equation}
    \int \frac{\text{d}^2 \mathbf{q}}{(2\pi)^2} e^{i\mathbf{q}\cdot \mathbf{x}} \int \frac{\text{d}^2\mathbf{l}}{(2\pi)^2} \frac{m^2}{[\mathbf{l}^2+m^2]^2[(\mathbf{l}+\mathbf{q})^2+m^2]} = \frac{mr}{8\pi^2 }K_0(mr)K_1(mr).
\end{equation}
The more complicated ``\Large$\tri{}$\normalsize"\;Fourier transform can be reduced to single integrals
\begin{align}
&\int \frac{\text{d}^2\mathbf{q}}{(2\pi)^2} \frac{e^{i\mathbf{q}\cdot \mathbf{x}}}{\mathbf{q}^2+m^2}\int\frac{\text{d}^2\mathbf{l}}{(2\pi)^2} \frac{1}{\left[\mathbf{l}^2+m^2\right][\left(\mathbf{l}+\mathbf{q}\right)^2+m^2]} \nonumber\\
&= \int \frac{\text{d}^2\mathbf{q}}{(2\pi)^2} \frac{\text{d}^2\mathbf{l}}{(2\pi)^2}\frac{\text{d}^2\mathbf{k}}{(2\pi)^2} \frac{e^{i\mathbf{q}\cdot \mathbf{x}}(2\pi)^2 \delta^{(2)}\left(\mathbf{k}-\mathbf{l}-\mathbf{q}\right)}{\left[\mathbf{l}^2+m^2\right][\mathbf{k}^2+m^2][\mathbf{q}^2+m^2]} \nonumber\\
&= \int \text{d}^2\mathbf{y}\frac{\text{d}^2\mathbf{q}}{(2\pi)^2} \frac{\text{d}^2\mathbf{l}}{(2\pi)^2}\frac{\text{d}^2\mathbf{k}}{(2\pi)^2} \frac{e^{-i\mathbf{l}\cdot \mathbf{y}}e^{i\mathbf{k}\cdot \mathbf{y}}e^{i\mathbf{q}\cdot (\mathbf{x}-\mathbf{y})}}{\left[\mathbf{l}^2+m^2\right][\mathbf{k}^2+m^2][\mathbf{q}^2+m^2]} \nonumber\\
&= \int \text{d}^2 \mathbf{y} \left[\int \frac{\text{d}^2\mathbf{q}}{(2\pi)^2}\frac{e^{i\mathbf{q}\cdot (\mathbf{x}-\mathbf{y})}}{\mathbf{q}^2+m^2}\right]\left[\int \frac{\text{d}^2\mathbf{l}}{(2\pi)^2}\frac{e^{-i\mathbf{l}\cdot \mathbf{y}}}{\mathbf{l}^2+m^2}\right]\left[\int \frac{\text{d}^2\mathbf{k}}{(2\pi)^2}\frac{e^{i\mathbf{k}\cdot \mathbf{y}}}{\mathbf{k}^2+m^2}\right]\nonumber\\
&= \left(\frac{1}{2\pi}\right)^3\int \text{d}^2 \mathbf{y} K_0\left(m|\mathbf{x}-\mathbf{y}|\right)K_0\left(m|\mathbf{y}|\right)^2 \nonumber\\
&= \left(\frac{1}{2\pi}\right)^3 \int_0^\infty \text{d}r'\; r' \; K_0\left(mr'\right)^2 \int_0^{2\pi} \text{d}\theta \; K_0\left(m\sqrt{r^2+r'^2-2rr' \cos(\theta)}\right) \nonumber\\
&= \frac{1}{4\pi^2} K_0\left(mr\right) \int_0^{mr} \text{d}r'\; r' \; K_0\left(mr'\right)^2 I_0\left(mr'\right)+\frac{1}{4\pi^2} I_0\left(mr\right) \int_{mr}^\infty \text{d}r'\; r' \; K_0\left(mr'\right)^3.
\end{align}
Here we have made use of the identity 
\begin{equation}
    \int_0^{2\pi}\text{d}\theta \cos(n\theta) K_0\left(m\sqrt{r^2+r'^2-2rr'\cos(\theta)}\right) = 
    \begin{cases}
        2\pi K_n(mr')I_n(mr) & r<r' \\
        2\pi K_n(mr)I_n(mr') & r>r'.
    \end{cases}
\end{equation}
We rewrite the first term using
\begin{equation}
    \int_0^{mr} \text{d}r'\; r' \; K_0\left(mr'\right)^2 I_0\left(mr'\right) = \frac{\pi}{3\sqrt{3}m^2} - \int_{mr}^\infty \text{d}r'\; r' \; K_0\left(mr'\right)^2 I_0\left(mr'\right),
\end{equation}
which gives 
\begin{equation}
    =\frac{K_0(mr)}{12\sqrt{3}\pi m^2} - \frac{K_0(mr)}{4\pi^2 m^2}\int_{mr}^\infty \text{d}\xi\; \xi \; K_0\left(\xi\right)^2 I_0\left(\xi\right)+\frac{I_0\left(mr\right)}{4\pi^2 m^2}  \int_{mr}^\infty \text{d}\xi\; \xi \; K_0\left(\xi\right)^3.
\end{equation}
We could leave the expression in this form, but to line-up more closely with the explicit expressions in \cite{deVega:1976xbp} we can use the simple Bessel function identities 
\begin{align}
    \int_{mr}^\infty \text{d}\xi\; \xi\; K_0\left(\xi\right)^2 I_0(\xi)  &= -mr K_0(mr)^2 I_1(mr) + 2\int_{mr}^\infty \text{d}\xi \;\xi \; K_0(\xi) K_1(\xi) I_1\left(\xi\right) \nonumber\\
    \int_{mr}^\infty \text{d}\xi\; \xi\; K_0\left(\xi\right)^3 &= mr K_0(mr)^2 K_1(mr) - 2\int_{mr}^\infty \text{d}\xi\; \xi \; K_0(\xi) K_1(\xi)^2,
\end{align}
to rewrite the above as
\begin{align}
    &\int \frac{\text{d}^2\mathbf{q}}{(2\pi)^2} \frac{e^{i\mathbf{q}\cdot \mathbf{x}}}{\mathbf{q}^2+m^2}\int\frac{\text{d}^2\mathbf{l}}{(2\pi)^2} \frac{1}{\left[\mathbf{l}^2+m^2\right][\left(\mathbf{l}+\mathbf{q}\right)^2+m^2]} \\
    &= \frac{K_0(mr)}{12\sqrt{3}\pi m^2} +\frac{K_0(mr)^2}{4\pi^2 m^2} \nonumber\\
    &\hspace{5mm}- \frac{K_0(mr)}{2\pi^2 m^2}\int_{mr}^\infty \text{d}\xi \;\xi \; K_0(\xi) K_1(\xi) I_1\left(\xi\right)-\frac{I_0\left(mr\right)}{2\pi^2 m^2}  \int_{mr}^\infty \text{d}\xi\; \xi \; K_0(\xi) K_1(\xi)^2.\nonumber
\end{align}

\bibliographystyle{JHEP}
\bibliography{cite.bib}
\end{document}